\documentclass[preprint,amsmath,amssymb,longbibliography]{revtex4-1}
\usepackage[english]{babel}
\usepackage{mathtools} 

\usepackage{color}

\usepackage{graphicx}
\usepackage{dcolumn}
\usepackage{bm}
\usepackage{hyperref}
\usepackage[mathlines]{lineno}

\usepackage{multirow}
\usepackage{booktabs}
\usepackage{enumitem}
\usepackage{float}

\newcommand{\dd}{\mathrm{d}}

\begin{document}

\preprint{APS/123-QED}

\title{ATP Level and Phosphorylation Free Energy Regulate Trigger-Wave Speed and Critical Nucleus Size in Cellular Biochemical Systems}

\author{Jianwei Li}
\affiliation{School of Physics, Center for Quantitative Biology, Peking University, Beijing 100871, China}

\author{Kai Meng}
\affiliation{School of Physics, Center for Quantitative Biology, Peking University, Beijing 100871, China}

\author{Xuewen Shen}
\affiliation{School of Physics, Center for Quantitative Biology, Peking University, Beijing 100871, China}

\author{Fangting Li}
\email{lft@pku.edu.cn}
\thanks{Corresponding author.}
\affiliation{School of Physics, Center for Quantitative Biology, Peking University, Beijing 100871, China}

\begin{abstract}

Trigger waves are self-regenerating propagating fronts that emerge from the coupling of nonlinear reaction kinetics and diffusion. In cells, trigger waves coordinate large-scale processes such as mitotic entry and stress responses. Although the roles of circuit topology and feedback architecture in generating bistability are well established, how nonequilibrium energetic driving shapes wave propagation is less well understood.
Here, we employ a thermodynamically consistent reaction--diffusion framework to investigate trigger-wave dynamics in ATP-dependent phosphorylation--dephosphorylation systems. We first recapitulate general expressions for trigger-wave speed in the bistable regime and analyze curvature-induced corrections that determine the minimum critical nucleus required for sustained propagation in higher dimensions. We then apply this framework to two representative systems, treating ATP concentration and the nonequilibrium parameter $\gamma = [ATP]/(K_{\mathrm{eq}}[ADP][P_i])$ as independent control variables to examine how energetic driving regulates wave propagation.
Our results show that ATP and $\gamma$ not only modulate wave speed, but can also reverse the direction of propagation and reshape the parameter regime supporting trigger waves. The critical excitation radius also depends on both ATP concentration and phosphorylation free energy. These findings identify the intracellular energetic state as a regulator of trigger-wave behavior, linking metabolic conditions to the spatial dynamics of wave propagation. More broadly, this framework connects classical reaction--diffusion theory with ATP-driven biochemical regulation and provides a general perspective on related energy-dependent cellular decision-making processes.

\end{abstract}

\maketitle

\title{ATP Level and Phosphorylation Free Energy Regulate Trigger Wave Speed and Determine the Critical Excitation Radius in Intracellular Biochemical Systems}

\section*{Introduction}

Nonequilibrium biochemical reactions can spontaneously generate stable and complex spatiotemporal structures, providing a foundation for modern nonlinear science and biophysics. The classical Belousov--Zhabotinsky (BZ) reaction demonstrated that, in open systems far from thermodynamic equilibrium, the homogeneous steady state is not the only attractor; instead, the system can generate structures such as rotating spirals and propagating wave fronts \cite{zaikinConcentrationWavePropagation1970,murray_mathematical_1989}. The propagation formula derived by Fisher in population dynamics further revealed a self-sustained propagation mechanism arising from the coupling of nonlinear reactions and diffusion \cite{fisherWaveAdvanceAdvantageous1937}. Subsequently, the reaction--diffusion model proposed by Turing showed that spatially heterogeneous structures can spontaneously emerge from a homogeneous background \cite{turingChemicalBasisMorphogenesis1952}. Studies of Turing-pattern stability and related nonlinear structures further clarified the principles of pattern selection in reaction--diffusion systems operating far from equilibrium \cite{ouyangTransitionUniformState1991,prigogine_introduction_1963}. Together, these works established a general theoretical framework for understanding how nonequilibrium reaction dynamics, coupled with diffusion, can generate propagating fronts and spatial patterns. These same principles are now recognized to underlie a wide range of biological phenomena, where cells and tissues exploit nonequilibrium chemical dynamics to produce functional spatiotemporal organization.

Beyond static spatial structures, biological systems often achieve rapid long-range information transmission through trigger waves at larger scales. A trigger wave is a self-regenerating propagating front arising from the coupling of local nonlinear reactions and diffusion, characterized by nearly constant amplitude and propagation speed during transmission \cite{gelensSpatialTriggerWaves2014}. A representative example is the mitotic trigger wave associated with CDK activation during division of Xenopus oocytes \cite{changMitoticTriggerWaves2013a}. Other examples include action potentials in axons \cite{hodgkinQuantitativeDescriptionMembrane1952} and calcium waves in muscle tissues \cite{allbrittonRangeMessengerAction1992}. Gelens and Ferrell systematically summarized the dynamical basis of trigger waves, pointing out that bistable or excitable dynamics constitute the key conditions for their emergence \cite{gelensSpatialTriggerWaves2014}. In Xenopus egg extract systems, trigger waves of Cdk1 activation were directly observed to propagate at an approximately constant speed, coordinating mitotic progression across millimeter-scale cytoplasm \cite{changMitoticTriggerWaves2013a}. Recent studies further show that the wave speed remains robust over a broad range of cytoplasmic concentrations, revealing that trigger waves can maintain stable dynamical properties even in complex cellular environments \cite{huangRobustTriggerWave2024}.

The generation of trigger waves relies on strong nonlinearities in the underlying reaction circuits, and such nonlinear behavior typically arises from positive feedback loops. During the cell-cycle G2--M transition, for example, activation of Cdk1/Cyclin B is controlled by Cdc25 and Wee1, forming an effectively coupled positive-feedback circuit \cite{pomereningBuildingCellCycle2003,shaHysteresisDrivesCellcycle2003a,novakDesignPrinciplesBiochemical2008}. Both theoretical and experimental studies have shown that this system can exhibit bistability and hysteresis over an appropriate parameter range, thereby conferring irreversibility and robustness on the cell-cycle transition \cite{pomereningBuildingCellCycle2003}. When coupled in space through diffusion, such local switch-like transitions can be extended into self-sustained propagating fronts. Bistable dynamics therefore provide the essential dynamical basis for mitotic trigger waves.

Importantly, bistability is not determined solely by circuit topology. Living systems continuously consume free energy to maintain steady states far from equilibrium. Phosphorylation–dephosphorylation (PdP) cycles use the free energy released from ATP hydrolysis to drive signal amplification and state transitions \cite{qianPhosphorylationEnergyHypothesis2007}. Although individual phosphorylation and dephosphorylation reactions are microscopically reversible, under non-equilibrium steady-state conditions the free-energy bias determined by the ATP/ADP/Pi ratio generates persistent flux and directionality in the cycle. The non-equilibrium steady-state theory developed by Qian and co-workers shows that the phosphorylation free energy $\Delta G = RT\ln\gamma$ can regulate both ultrasensitivity and the existence of bistable regimes \cite{qianPhosphorylationEnergyHypothesis2007}. Therefore, bistability and hysteresis in PdP cycles are fundamentally dynamical structures maintained by energy dissipation rather than purely by feedback topology. Experiments on the fission yeast G2–M transition further demonstrate that ATP levels can influence the bistability and hysteresis of Cdc2 activation \cite{zhaoNonequilibriumNonlinearKinetics2016}. Meanwhile, single-cell ATP imaging studies have revealed substantial spatiotemporal heterogeneity in intracellular ATP concentrations \cite{yaginumaDiversityATPConcentrations2014}, suggesting that the cellular energetic state may act as an important control parameters for spatiotemporal propagation behavior.

Motivated by these considerations, we develop a unified theoretical framework that incorporates spatial diffusion and nonequilibrium energy supply into bistable dynamics. Under thermodynamically consistent conditions, we begin with a general bistable reaction term, recapitulate the expressions for trigger-wave speed and their curvature corrections, and further examine the minimum triggering nucleus radius in higher-dimensional systems. Although related wave forms have been discussed in reaction--diffusion theory and pattern-forming systems, their systematic treatment in the context of energy-driven bistable trigger waves remains incomplete. To address this gap, we treat ATP concentration and the free-energy parameter $\gamma$, which characterizes the distance from equilibrium, as two independent control parameters, and investigate how they regulate propagation direction, the existence of traveling fronts, and the minimum trigger-nucleus size.

To connect the theoretical analysis with specific biological systems, we consider two representative reaction circuits. The first is the autophosphorylation circuit of Rad53, the budding-yeast checkpoint kinase involved in the S-phase DNA damage response \cite{pellicioliActivationRad53Kinase1999}, which serves as a minimal model of a kinase positive-feedback system. The second is the mitotic trigger-wave system associated with Cdk1 activation during the G2--M transition \cite{changMitoticTriggerWaves2013a}, which has been extensively studied in vitro. By explicitly introducing energy-related parameters into these models, we derive several experimentally testable predictions, including how wave speed depends on ATP concentration, the conditions under which reverse propagation may occur, and the minimum trigger-nucleus size required for sustained propagation in two- and three-dimensional spaces.

Overall, this work aims to establish a quantitative link between classical reaction--diffusion theory and nonequilibrium biochemistry. Rather than focusing solely on circuit topology or empirical fitting, we emphasize the roles of ATP concentration and the free energy of ATP hydrolysis, which together characterize the energetic driving and dissipation of the system, in determining the properties of spatiotemporal propagation. Treating these energetic quantities as dynamical control parameters provides a theoretical framework that may help explain how metabolic conditions influence spatiotemporal propagation in ATP-driven cellular decision-making processes, and may also offer a more general perspective on wave propagation in related bistable systems.

\section{Trigger-wave speed jointly determined by biochemical reactions and spatial diffusion}

To clarify the physical picture of trigger waves and to establish the theoretical basis for subsequent discussions, we briefly summarize key mathematical results for trigger waves in bistable reaction--diffusion systems. In this work, the term \emph{trigger wave} specifically refers to waves arising from dynamical bistability; they correspond to the propagation of a wavefront connecting the high and low stable states in space, thereby driving interconversion between the two states.

\subsubsection{Propagation direction is determined by the potential difference $\Delta F$}

\begin{figure}[h!]
\centering
\includegraphics[width=0.7\linewidth]{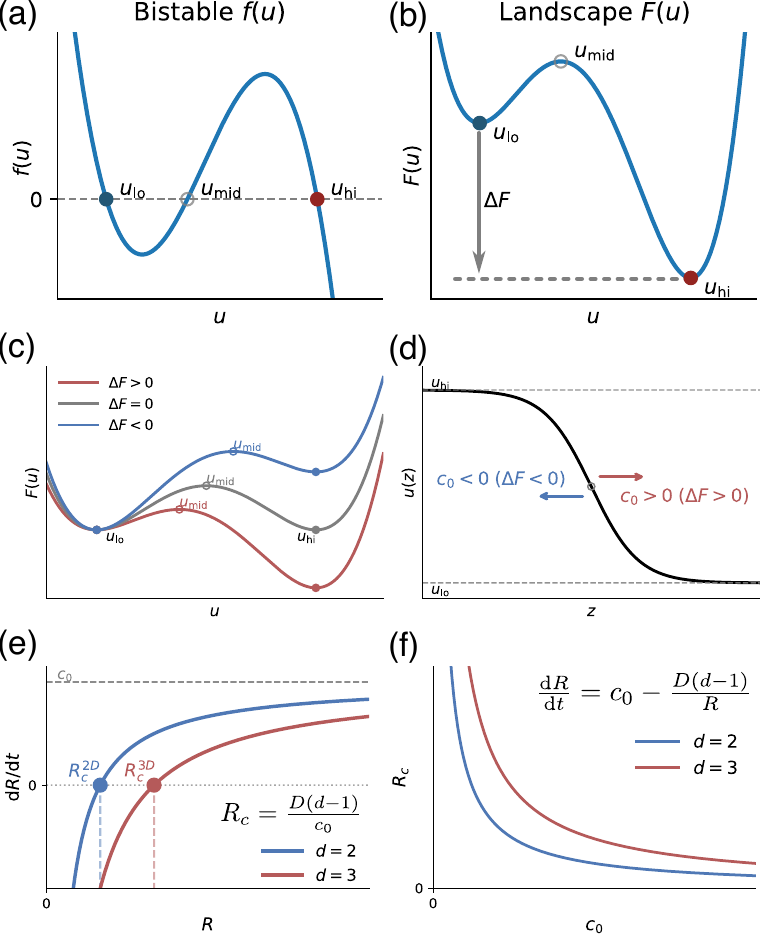}
\caption{
Schematic illustration of bistable reaction kinetics and trigger-wave propagation.
(a) Reaction term $f(u)$ versus $u$. Three real roots of $f(u)=0$ correspond to two stable steady states, $u_{\mathrm{lo}}$ and $u_{\mathrm{hi}}$, separated by an unstable state $u_{\mathrm{mid}}$.
(b) Potential landscape defined by $F(u)=-\int^u f(u')\,\mathrm{d}u'$, with two wells for the stable states and a barrier for the unstable state.
(c) Potential landscapes with identical local bistability but different global biases, corresponding to $\Delta F>0$, $\Delta F=0$, and $\Delta F<0$. The sign of $\Delta F$ sets the preferred invasion direction.
(d) Traveling-front profile $u(z)$. The sign of $\Delta F$ determines the propagation direction, and the wave speed follows a Luther-type scaling, $c_0\sim\sqrt{Dk_{\mathrm{rxn}}}$.
(e--f) Curvature correction and critical nucleus for spherical trigger waves in higher dimensions. The radial speed obeys $\mathrm{d}R/\mathrm{d}t=c_0-(d-1)D/R$, where $d$ is the spatial dimension. Fronts with $R<R_c$ collapse, whereas those with $R>R_c$ expand into sustained trigger waves.
}
\label{fig:bistable_fu_DeltaF}
\end{figure}

Consider a single-variable reaction--diffusion equation for $u$,
\begin{equation}
\frac{\partial u}{\partial t}
=
D \nabla^2 u + f(u,\theta),
\label{eq:RD_general}
\end{equation}
where $u(\bm{x},t)$ denotes the concentration (or activation fraction) of an activated molecular species, $D$ is an effective diffusion coefficient, $\theta$ is a set of control parameters, and $f(u,\theta)$ is the local reaction-kinetics term. Bistability means that, for fixed parameters, the system admits two stable fixed points $u_{\mathrm{lo}}$ and $u_{\mathrm{hi}}$, separated by an unstable fixed point $u_{\mathrm{mid}}$. A schematic illustration is shown in Fig.~\ref{fig:bistable_fu_DeltaF}(a).

Given that $f$ depends only on the single variable $u$, we define a potential-like function $F(u;\theta)$ by
\begin{equation}
\frac{\partial F}{\partial u}(u;\theta) = -f(u;\theta),
\label{eq:potential_def}
\end{equation}
so that the reaction term can be interpreted as the negative gradient of an effective potential.

The ``potential difference'' between the two stable states, i.e., the effective driving force, can be written as
\begin{equation}
\Delta F(\theta)
\equiv
F(u_{\mathrm{lo}};\theta)-F(u_{\mathrm{hi}};\theta)
\equiv
\int_{u_{\mathrm{lo}}}^{u_{\mathrm{hi}}} f(u;\theta)\,\dd u .
\label{eq:DeltaF_def}
\end{equation}

As illustrated in Fig.~\ref{fig:bistable_fu_DeltaF}(b), the potential difference $\Delta F$ characterizes the thermodynamic tendency of the system to evolve from the higher-potential stable state toward the lower-potential one. When $\Delta F>0$, the high-activity state becomes globally preferred under nonequilibrium driving; when $\Delta F<0$, the low-activity state becomes globally preferred; and $\Delta F=0$ corresponds to the Maxwell point at which the two states are energetically equivalent.

For such bistable systems, the planar trigger-wave speed $c_0$ satisfies the general identity
\begin{equation}
c_0
=
\frac{\Delta F(\theta)}
{\displaystyle \int_{-\infty}^{+\infty}
\left(\frac{\dd u}{\dd z}\right)^2 \dd z},
\label{eq:c_ratio_form}
\end{equation}
where $\dd u/\dd z$ is the spatial slope of the wave front. A detailed derivation is provided in Appendix~\ref{appendix:wave_speed_derivation}.

Since any nontrivial monotonic front satisfies
\begin{equation}
\int_{-\infty}^{+\infty}
\left(\frac{\dd u}{\dd z}\right)^2 \dd z > 0,
\end{equation}
the sign of $c_0$ is determined solely by the sign of $\Delta F$. As shown schematically in Fig.~\ref{fig:bistable_fu_DeltaF}(c--d), if $\Delta F>0$ then $c_0>0$ and the front propagates forward, with the high state invading the low state; if $\Delta F<0$ the front propagates backward and the high state is engulfed by the low state; if $\Delta F=0$ then $c_0=0$, corresponding to a stationary interface in one dimension.

Equation~\eqref{eq:c_ratio_form} shows that $\Delta F(\theta)$ appears in the numerator and determines the propagation direction, while the speed magnitude is also influenced by the denominator (the integral term). In traveling-wave propagation, the front advances with speed $c_0$. Let $D$ denote the diffusion coefficient and $k_{\mathrm{rxn}}$ the effective rate constant of the underlying positive-feedback reaction. If the front completes a local switching event over a characteristic reaction timescale $\tau_{\mathrm{rxn}}$ and advances over a characteristic length scale $\ell$ during that time, then $c_0 \sim \ell/\tau_{\mathrm{rxn}}$. Using the estimates $\ell \sim \sqrt{D\tau_{\mathrm{rxn}}}$ and $\tau_{\mathrm{rxn}} \sim 1/k_{\mathrm{rxn}}$, one obtains
\begin{equation}
c_0 \sim \sqrt{D\,k_{\mathrm{rxn}}}.
\label{eq:c_scale_rxn}
\end{equation}
This scaling estimate is the Luther relation \cite{murray_mathematical_1989}. It is not an exact result, but a leading-order scaling approximation for the magnitude of the wave speed. In particular, it does not determine the propagation direction and is reliable only away from the near-balanced regime $\Delta F \approx 0$.

\subsubsection{Curvature correction to trigger-wave propagation}

In three spatial dimensions ($d=3$), trigger-wave fronts typically take the form of curved spherical interfaces. When the intrinsic front thickness $\ell$ is much smaller than the radius of curvature $R$, a thin-front approximation applies and the front may be treated as a geometric interface evolving in time. Consider the radially symmetric case $u(\mathbf{x},t)=u(r,t)$ with $r=|\mathbf{x}|$. In $d$-dimensional radial coordinates, the reaction--diffusion equation leads to
\begin{equation}
\frac{\dd R}{\dd t}
=
c_0-\frac{D(d-1)}{R},
\label{eq:Rdot_curvature}
\end{equation}
showing that the normal propagation speed of a spherical front receives a first-order correction proportional to the curvature $\kappa=(d-1)/R$ relative to the planar wave speed $c_0$. A detailed derivation is given in Appendix~\ref{appendix:curvature_correction}. In the limit $R\to\infty$, the curvature vanishes, and the normal speed of the spherical front approaches the one-dimensional planar trigger-wave speed $c_0$. Equation~\eqref{eq:Rdot_curvature} is the corresponding eikonal relation for reaction--diffusion systems, $v_n=c_0-D\kappa$. Figure~\ref{fig:bistable_fu_DeltaF}(e--f) schematically illustrates the radial dynamics of a spherically symmetric trigger nucleus.

For small $R$, the curvature term dominates, so that $\dd R/\dd t<0$ and the trigger nucleus collapses as the high-activity region shrinks. Setting $\dd R/\dd t=0$ gives the critical nucleus radius
\begin{equation}
R_c=\frac{D(d-1)}{c_0}.
\label{eq:Rc_general}
\end{equation}
When $R>R_c$, the front expands stably outward; as $R$ increases further, the propagation speed asymptotically approaches the planar-wave limit.

\section{Trigger waves in Rad53 activation in the S-phase checkpoint}

IIn eukaryotic cells, the DNA damage checkpoint constitutes one of the core regulatory mechanisms governing cell-cycle control. Its basic architecture typically involves damage-sensing kinases, signal-transducing kinases, and downstream effector molecules \cite{hartwellCheckpointsControlsThat1989,morganCellCyclePrinciples2007}. Through self-reinforcing phosphorylation reactions, this circuitry amplifies and sustains checkpoint signaling and, when necessary, arrests cell-cycle progression to safeguard genome stability \cite{elledgeCellCycleCheckpoints1996,bartekDNADamageCheckpoints2007,yatesDNADamageReplication2025}. This regulatory architecture is highly conserved across eukaryotes: from yeast to mammals, checkpoint signaling follows a broadly similar hierarchical kinase structure \cite{nurseUniversalControlMechanism1990a,zhouDNADamageResponse2000,harperDNADamageResponse2007a}.

In budding yeast \textit{Saccharomyces cerevisiae}, Rad53 is the central effector kinase of the DNA damage checkpoint \cite{sanchezRegulationRAD53ATMlike1996,pellicioliActivationRad53Kinase1999}. In the absence of DNA damage, Rad53 predominantly remains in a low-phosphorylation, low-activity state; upon damage, upstream sensing modules centered on Mec1/Tel1 are activated and initiate Rad53 phosphorylation. Phosphorylated Rad53 can dimerize via FHA-domain--mediated interactions, which enhances trans-phosphorylation of inactive Rad53 and establishes a positive-feedback amplification loop \cite{gilbertBuddingYeastRad92001,smolkaProteomewideIdentificationVivo2007}. This positive-feedback topology introduces nonlinearity and, under appropriate conditions, can give rise to bistable switch-like behavior.

From a thermodynamic perspective, Rad53 phosphorylation is coupled to ATP hydrolysis. A complete activation--deactivation cycle consumes ATP and releases the associated hydrolysis free energy,
\[
\Delta G = k_B T \ln \gamma,
\qquad
\gamma = K_{eq} \frac{[\mathrm{ATP}]}{[\mathrm{ADP}][\mathrm{Pi}]},
\]
where $K_{eq}$ is the equilibrium constant for ATP hydrolysis. Consequently, the checkpoint circuit is intrinsically a nonequilibrium steady-state reaction system maintained by continuous energy input \cite{qianPhosphorylationEnergyHypothesis2007, tuNonequilibriumMechanismUltrasensitivity2008}. Under appropriate approximations, its reaction kinetics can be reduced to an effective equation containing cubic nonlinearity \cite{jinNonequilibriumStochasticityInfluence2018}. It should be emphasized that, due to the small cell size of budding yeast, the Rad53 circuit does not form an experimentally observable spatial trigger wave in intact cells. Here we nevertheless adopt the Rad53 notation and regulatory mechanism as a minimal representative model of a PdP cycle with positive feedback, in order to discuss the trigger-wave dynamics that energy-driven bistable biochemical systems may exhibit under spatial coupling.

\subsection{Regulation of trigger-wave speed and critical radius in Rad53 activation by ATP and hydrolysis free energy}

\begin{figure}[h!]
\centering
\includegraphics[width=0.8\linewidth]{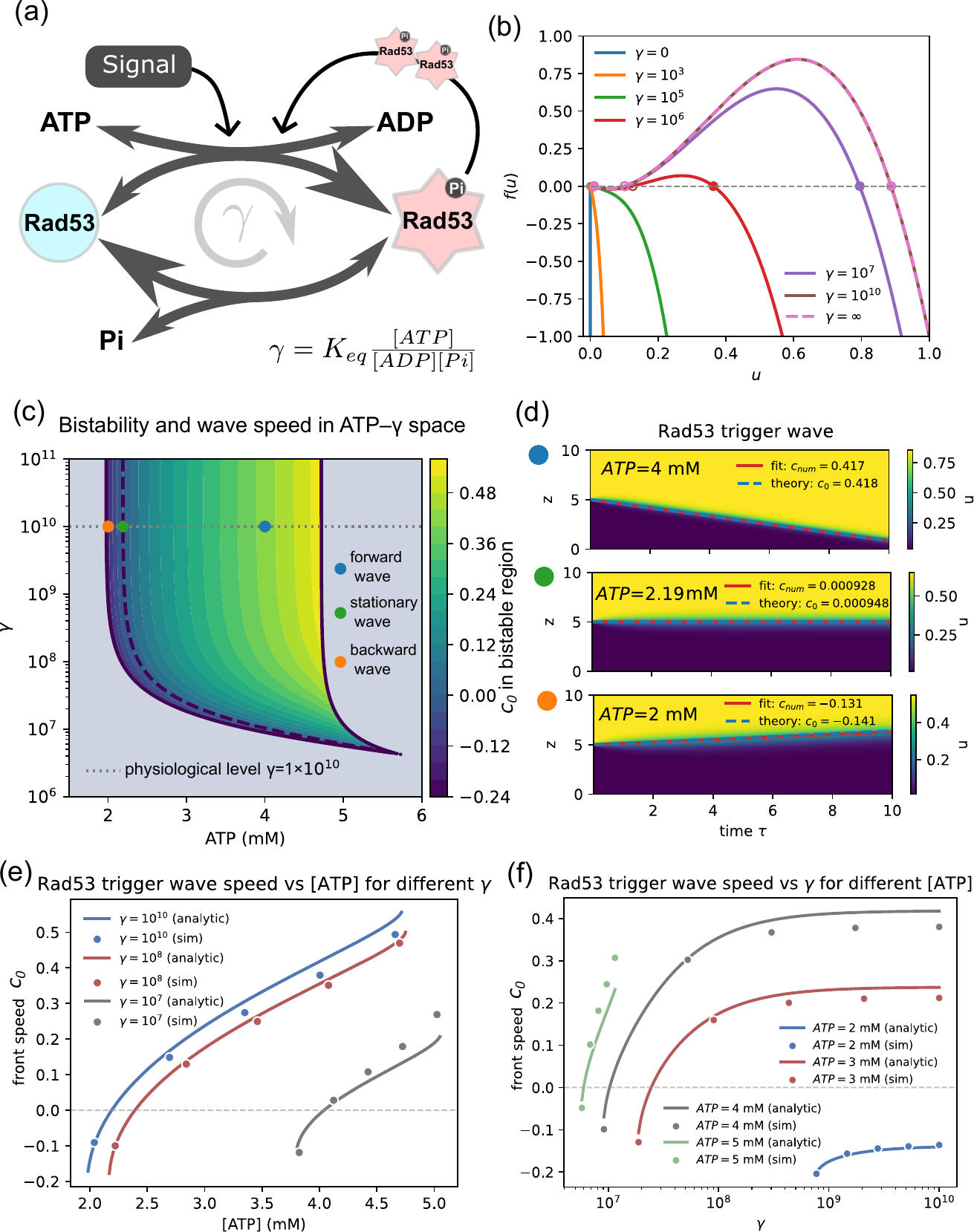}
\caption{
ATP and hydrolysis free energy regulate trigger waves in S-phase checkpoint Rad53 activation.
(a) Rad53 activation circuit coupled to ATP hydrolysis.
(b) Fixed points of the local reaction term $f(u)$ for varying non-equilibrium drive $\gamma$.
(c--d) Bistable region and planar trigger-wave speed $c_0$ in the $\mathrm{ATP}$--$\gamma$ phase plane; representative parameter points (three cases) are selected and highlighted for the example kymographs in (d).
(e--f) Trigger-wave speed versus $[\mathrm{ATP}]$ and $\gamma$; symbols are PDE simulations, and solid lines are theoretical predictions from the analytical wave-speed formula.
}
\label{fig:Rad53_scheme}
\end{figure}

Figure~\ref{fig:Rad53_scheme}(a) illustrates the activation circuit and its coupling to ATP hydrolysis. Based on a thermodynamically consistent derivation (see Appendix~\ref{app:Rad53_thermo_derivation}), the reaction term can be written as the difference between the far-from-equilibrium limit and a finite nonequilibrium correction,
\begin{equation}
f_{\mathrm{full}}(u)
=
f_{\infty}(u)
-
\varepsilon\,g(u),
\qquad
\varepsilon=\frac{1}{\gamma},
\label{eq:f_full_decomp_gamma}
\end{equation}
where, in the limit $\gamma\to\infty$ ($\varepsilon\to 0$), the reaction term reduces to
\begin{equation}
f_{\infty}(u)
=
\frac{1}{\alpha}(1-u)
+
\frac{[ATP]}{\alpha}\Bigl(\xi_0+\eta_0 u^2\Bigr)(1-u)
-
u,
\label{eq:f_infty_explicit}
\end{equation}
and
\begin{equation}
g(u)
=
[ATP]\Bigl(\xi_0+\eta_0 u^2\Bigr)u.
\label{eq:g_u_def}
\end{equation}

Here $\xi_0$ encodes the strength of the upstream DNA-damage input (e.g., via Mec1/Tel1-mediated priming), whereas $\eta_0$ is a damage-independent parameter that sets the intrinsic nonlinear feedback strength.

Figure~\ref{fig:Rad53_scheme}(b) shows how the shape of the reaction term $f(u)$ evolves with $\gamma$. As $\gamma$ decreases, the effective driving force at large $u$ is progressively suppressed, leading to the merger and eventual disappearance of fixed points.

Within the bistable regime, since the local kinetics reduce to a cubic form, the planar wave speed can be obtained analytically (see Appendix~\ref{app:Rad53_ATP_wave_speed_derivation}), yielding
\begin{equation}
c_0
\sim
\sqrt{
\frac{D k_-}{2}
\cdot
\frac{\eta_0}{\alpha}
\,[ATP]
}
\left(
r_{\mathrm{lo}}+r_{\mathrm{hi}}-2r_{\mathrm{mid}}
\right).
\label{eq:c0_atp_scaling}
\end{equation}
Therefore, in parameter ranges where the root-combination term varies moderately, the dominant scaling is $c_0 \propto \sqrt{[ATP]}$. Substituting into the critical-radius formula~\eqref{eq:Rc_general} further gives $R_c \propto 1/\sqrt{[ATP]}$ under the same approximation. This result provides a clear, experimentally testable prediction: increasing $[\mathrm{ATP}]$ accelerates trigger waves and reduces the minimal trigger-nucleus size, thereby facilitating spatial propagation.

In numerical simulations, we adopted the same parameter scaling as in Dianjieli \textit{et al.} \cite{liInterplayATPHydrolysis2024a} and set the key dimensionless parameter to $\alpha = 10^{7}$. Figure~\ref{fig:Rad53_scheme}(c) summarizes the global dynamics of the Rad53 system in the $\mathrm{ATP}$--$\gamma$ phase plane. Within the bistable region, we evaluated the stable planar trigger-wave speed $c_0$ using the analytical expression derived above and visualized it with a continuous color scale. The dashed curve marks the condition $\Delta F=0$, where the two stable states are energetically equivalent and thus $c_0 \to 0$. This curve separates two propagation regimes: on the left (lower $[\mathrm{ATP}]$ and smaller $\gamma$), $\Delta F>0$ and the system supports reverse waves, with the high-activity state retreating into the low state; on the right (higher $[\mathrm{ATP}]$ and stronger nonequilibrium driving), $\Delta F<0$ and the high-activity state invades the low state as a stable forward-propagating trigger wave.

To validate this phase-diagram structure, we selected three representative parameter sets within physiologically relevant $\gamma$ ranges, located in the reverse-wave region, near the stationary boundary, and in the forward-wave region, respectively, and marked them by different colors in Fig.~\ref{fig:Rad53_scheme}(c). The corresponding one-dimensional PDE simulations and analytical wave-speed predictions are compared in Fig.~\ref{fig:Rad53_scheme}(d). Both the propagation direction and the magnitude of the wave speed show excellent agreement between numerical results and analytical theory, indicating that the wave-front analysis provides reliable quantitative predictions across the bistable parameter regime.

Furthermore, fixing $\gamma$ and varying $[\mathrm{ATP}]$ alone yields the trends in Fig.~\ref{fig:Rad53_scheme}(e), where colors indicate different $\gamma$. The PDE simulation data (symbols) closely follow the analytical prediction (solid lines) and obey the dominant scaling in Eq.~\eqref{eq:c0_atp_scaling}, giving a monotonic increase of $c_0$ with $[\mathrm{ATP}]$ within the forward-wave regime. Beyond this leading scaling, the propagation direction is set by the sign of $\Delta F$ (determined by the three fixed points), which yields $c_0\approx 0$ near $\Delta F=0$ and negative (reverse) speeds at low $[\mathrm{ATP}]$. As $\gamma$ decreases, the overall magnitude of $c_0$ is suppressed; simultaneously, the bistable window shifts to higher $[\mathrm{ATP}]$ and progressively shrinks until it vanishes, consistent with the phase-diagram structure in Fig.~\ref{fig:Rad53_scheme}(c). 

Similarly, Fig.~\ref{fig:Rad53_scheme}(f) shows the dependence of wave speed on $\gamma$ at fixed ATP. For large $\gamma$, the wave speed approaches a plateau primarily set by ATP level, reflecting that in the strongly nonequilibrium limit ATP becomes the dominant rate-limiting factor. For small $\gamma$, the bistable structure collapses and the system can no longer sustain forward trigger waves, exhibiting only decay or reverse propagation.

\subsection{Spherical trigger-wave propagation jointly determined by ATP hydrolysis free energy and curvature}

\begin{figure}[h!]
\centering
\includegraphics[width=1\linewidth]{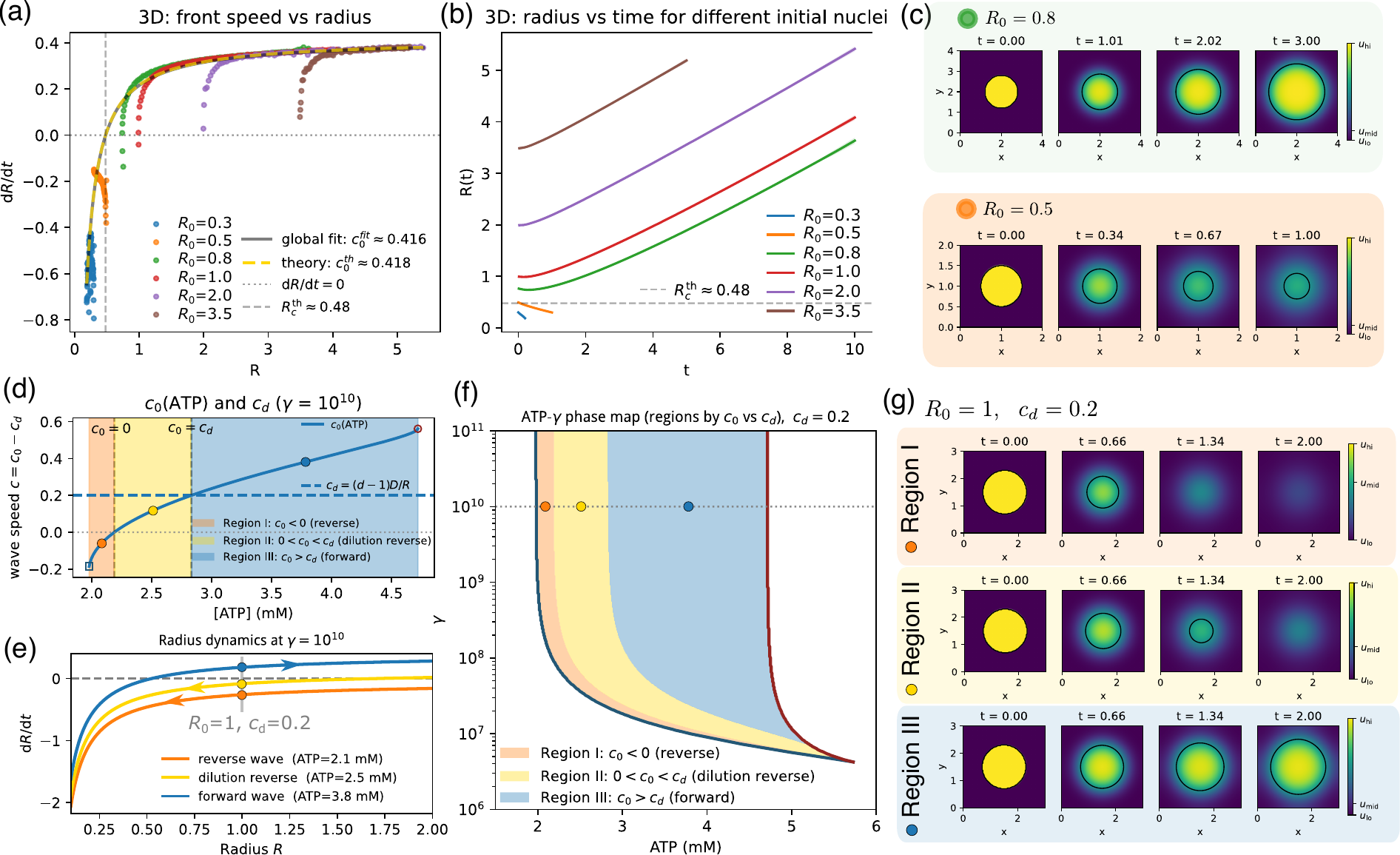}
\caption{
Curvature correction and critical radius for spherical trigger waves.
(a) Radial speed $\dd R/\dd t$ as a function of nucleus radius $R$.
(b) Radius dynamics $R(t)$ for different initial radii.
(c) Spatiotemporal evolution of the three-dimensional reaction--diffusion field for different initial radii.
(d--f) ATP dependence of the planar wave speed $c_0$ and the curvature-induced correction $c_d$, showing three propagation regimes.
(i) Region I: reverse-trigger regime, where $c_0<0$ and the high-activity region shrinks regardless of the initial radius.
(ii) Region II: dilution-dominated regime, where $0<c_0<c_d$ and the initial high-activity nucleus is lost because curvature-induced dilution prevents expansion.
(iii) Region III: forward-trigger regime, where $c_0>c_d$ and the nucleus expands into a sustained trigger wave.
(g) Three-dimensional simulations at fixed initial radius with different ATP levels corresponding to the three regimes.
In panels (c) and (g), the displayed $x$--$y$ sections pass through the equatorial plane of the spherical nucleus.
}
\label{fig:3D_dynamics}
\end{figure}

Beyond the planar, effectively one-dimensional setting, front curvature must be taken into account. In three dimensions, Figs.~\ref{fig:3D_dynamics}(a,b) show that, after the transient relaxation of the initially step-like profile, fronts initialized with different radii $R_0$ converge to the theoretical relation
\[
\frac{\dd R}{\dd t}
=
c_0-\frac{2D}{R},
\]
and the value of $c_0$ extracted from PDE simulations agrees with the analytically computed planar trigger-wave speed. The evolution $R(t)$ further shows that only when $R_0$ exceeds a larger three-dimensional critical radius $R_c^{(3D)}$ can the high-activation region expand persistently.

The full three-dimensional spatiotemporal evolution (Fig.~\ref{fig:3D_dynamics}(c)) provides a direct visualization of this behavior: for sufficiently large $R_0$, the high-activity state forms a steadily expanding spherical front; for $R_0<R_c^{(3D)}$, the initially activated region is diluted by diffusion, becomes diffuse, and gradually disappears. This confirms that trigger-wave formation depends not only on the planar limiting speed set by metabolic driving, but also on spatial geometry.

Even when metabolic parameters are fixed and the planar wave speed satisfies $c_0>0$, curvature can still suppress expansion through diffusion-driven dilution, introducing a critical radius jointly determined by geometric dimension and nucleus size. To describe curvature-induced spatial corrections across dimensions within a unified framework, and to compare them directly with the metabolically determined planar speed $c_0(\mathrm{ATP},\gamma)$, we introduce a dilution term with units of velocity,
\begin{equation}
c_d \equiv \frac{D(d-1)}{R},
\label{eq:cd_def}
\end{equation}
where $d$ is the spatial dimension and $R$ is the instantaneous nucleus radius. Under the thin-front approximation, $c_d$ quantifies the effective speed reduction caused by curvature-driven spatial dilution in higher dimensions, and is precisely the curvature term in the eikonal relation. Using the curvature-corrected propagation law~\eqref{eq:Rdot_curvature}, the radial dynamics of a spherically symmetric front can be written compactly as
\begin{equation}
\frac{\dd R}{\dd t}
=
c_0(\mathrm{ATP},\gamma)-c_d(R),
\label{eq:Rdot_cd}
\end{equation}
so that the criterion for nucleus expansion reduces to comparing the relative magnitudes of $c_0$ and $c_d$.

Within the physiological range $\gamma=10^{10}$, the planar wave speed $c_0(\mathrm{ATP},\gamma)$ increases monotonically with ATP and changes propagation direction at $\Delta F=0$. In contrast, for a fixed initial radius $R_0$, the spatial term $c_d$ can be treated as an approximately constant scale. As shown in Fig.~\ref{fig:3D_dynamics}(d), comparing $c_0$ and $c_d$ partitions the propagation behavior into three regimes: the metabolic reverse-wave regime (Region I) with $c_0<0$, where propagation is backward even in the planar limit (Fig.~\ref{fig:3D_dynamics}(e)) due to the potential bias $\Delta F>0$; the dilution-dominated regime (Region II) with $0<c_0<c_d$, where the planar system would support forward propagation but the nucleus is too small, so diffusion-driven dilution causes the activated region to decay and become diffuse; and the forward-trigger regime (Region III) with $c_0>c_d$, where metabolic driving overcomes spatial dilution and the nucleus expands into a sustained trigger wave. Therefore, the emergence of forward trigger waves depends not only on $\Delta F$ (which determines the sign of $c_0$) but also on spatial scale through the additional threshold introduced by $c_d$.

Extending this classification to the $\mathrm{ATP}$--$\gamma$ phase plane, we further subdivide the bistable region into three propagation regimes, as shown in Fig.~\ref{fig:3D_dynamics}(f) (example shown for $c_d=0.2$). The dark blue and dark red curves mark the left and right boundaries of the bistable region, respectively, while the background partition is determined by the comparison between $c_0$ and $c_d$: Region I corresponds to $c_0<0$, Region II to $0<c_0<c_d$, and Region III to $c_0>c_d$. This quantitative partition can be regarded as a natural extension of Fig.~\ref{fig:Rad53_scheme}(c).

To validate the metabolic--spatial criterion above, we selected one representative parameter set from each region and performed numerical simulations of the three-dimensional reaction--diffusion equation. The results are shown in Fig.~\ref{fig:3D_dynamics}(g). For $c_d=0.2$ (corresponding to $R_0=1$ in the three-dimensional simulations), the numerical results reproduce the tripartite classification: in Region I, $c_0<0$ and the activated region decays, corresponding to a tendency toward reverse propagation; in Region II, although $c_0>0$, the condition $c_0<c_d$ prevents expansion and the nucleus shrinks away; in Region III, where $c_0>c_d$, the high-activity region expands and forms a forward trigger wave. These results indicate that ATP and $\gamma$ jointly regulate trigger-wave propagability together with spatial curvature: $c_0$ sets the maximal propagation capacity in the planar limit, whereas $c_d$ represents the geometric suppression induced by curvature.

The curvature effect for spherical fronts may be testable in reconstituted \emph{in vitro} systems by generating a localized high-activity region. In a quasi-two-dimensional geometry, an approximately two-dimensional reaction environment could be formed by confining cell extract between two glass slides and initializing the system in the low-activity steady state. Local injection of high-activity extract would then create a nucleus with a controlled initial radius. The theory suggests two possible outcomes: nuclei above a critical size may expand into outward-propagating trigger waves, whereas smaller nuclei may decay because of dilution by the surrounding low state. The model also suggests that higher ATP levels would reduce the critical nucleus size and increase the wave speed. A similar strategy could be extended to three-dimensional settings by preparing an initial spherical high-activity nucleus in a low-state background and monitoring its subsequent evolution.


\section{Trigger waves in CDK activation}
\label{sec:g2m_triggerwave_background}

In the cell cycle, the G2--M transition marks the entry of a cell from a post-replication preparatory state into the highly ordered yet intrinsically high-risk process of mitosis. As the final regulatory checkpoint before mitotic entry, the G2--M checkpoint must respond to mitosis-promoting signals and coordinate the spatially synchronized onset of division throughout the cell. Because mitotic entry is effectively irreversible once initiated, this transition must be executed in a rapid, coordinated, and reliable manner.

At the molecular level, the core of the G2--M transition is activation of cyclin-dependent kinase (CDK). This regulatory architecture is evolutionarily highly conserved, following the same basic logic from fission yeast to vertebrates\cite{morganCellCyclePrinciples2007,perryCdc25Wee1Analogous2007,lemonnierG2toMTransitionPhosphatase2020}. In fission yeast \textit{Schizosaccharomyces pombe}, the CDK Cdc2 forms a complex with the B-type cyclin Cdc13, which is synthesized continuously during G2; activation of this complex is promoted by the phosphatase Cdc25 and inhibited by the kinase Wee1 \cite{nurseUniversalControlMechanism1990a}. In higher eukaryotes, the analogous module consists of cyclin~B--Cdk1, Cdc25, and Wee1/Myt1, and has been systematically validated in \textit{Xenopus laevis} egg extracts, \textit{Drosophila}, and mammalian cells \cite{ferrellCellCycleTyrosine1991, pomereningBuildingCellCycle2003, morganCellCyclePrinciples2007}.

From the perspective of circuit architecture (Fig.~\ref{fig:g2m_f_and_F}(a)), the regulatory core of the G2--M transition is built around coupled positive and double-negative feedback loops centered on the Cdc13/Cdc2 complex (here referred to as the CDK complex). Active CDK phosphorylates Cdc25 at multiple sites, thereby enhancing its phosphatase activity; activated Cdc25 in turn dephosphorylates and activates CDK, forming a self-reinforcing positive feedback loop. In parallel, CDK phosphorylates and inhibits Wee1, and the resulting reduction in Wee1 activity relieves the inhibitory phosphorylation of CDK, thereby generating a double-negative feedback loop. Coupling of these loops through the central CDK complex, together with multisite phosphorylation--dephosphorylation cycles, introduces nonlinearity into the regulatory response, allowing CDK activity to exhibit bistability with respect to inputs such as cyclin concentration.

These phosphorylation cycles are intrinsically nonequilibrium processes continuously driven by the free energy released from ATP hydrolysis. This hydrolysis free energy provides the energetic basis for rapid, synchronized, and effectively irreversible switching during the G2--M transition, and thereby helps shape its dynamical properties.

In this reaction circuit, the total concentration of the Cdc13/Cdc2 complex, $[\mathrm{Cdc13/Cdc2}^T]$, including both active and inactive forms---plays two roles. First, it enters the reaction kinetics and modifies the effective reaction term $f$. Second, it accumulates progressively during G2 and thereby acts as a triggering signal for mitotic entry. Accordingly, in what follows, we focus on three control variables: $[\mathrm{Cdc13/Cdc2}^T]$, ATP, and $\Delta G$. We first analyze an idealized system without APC-mediated Cdc13 degradation, and then extend the model to include cyclin synthesis and accumulation together with APC-mediated cyclin degradation. Within this framework, we examine the existence, direction, and speed of trigger waves under different energetic and biochemical conditions, and derive experimentally testable predictions relevant to \emph{in vitro} systems.

\subsection{Idealized system with fixed $[\mathrm{Cdc13/Cdc2}^T]$ and no APC-mediated Cdc13 degradation}
\subsubsection{Effect of ATP hydrolysis free energy on Cdk trigger waves}

We first treat the total CDK complex concentration $[{\rm Cdc13/Cdc2}^T]$ (denoted $T$) as an external parameter and keep it fixed, while neglecting APC-mediated degradation/reset processes. Under this assumption, the dynamics can be reduced to a single-variable form\cite{zhaoNonequilibriumNonlinearKinetics2016}:
\begin{equation}
\frac{\mathrm d A}{\mathrm d t} = J_P - J_W \equiv f(A;T,[\mathrm{ATP}],\gamma),
\label{eq:g2m_fixedT_local_ode}
\end{equation}
where $A$ denotes the concentration (in nM) of the active complex (i.e., $[\mathrm{Cdc13/Cdc2}^A]$), and $J_P$ and $J_W$ represent the forward and reverse fluxes associated with the Wee1- and Cdc25-mediated branches of the PdP cycle. Because both Wee1 and Cdc25 activities are regulated by $A$, and assuming rapid equilibration of these regulators\cite{zhaoNonequilibriumNonlinearKinetics2016}, the effective reaction term can be expressed as a function of ATP, $\gamma$, and $T$ (see Appendix~\ref{appendix:g2m_model} for details).

\begin{figure}[h]
  \centering
  \includegraphics[width=1\linewidth]{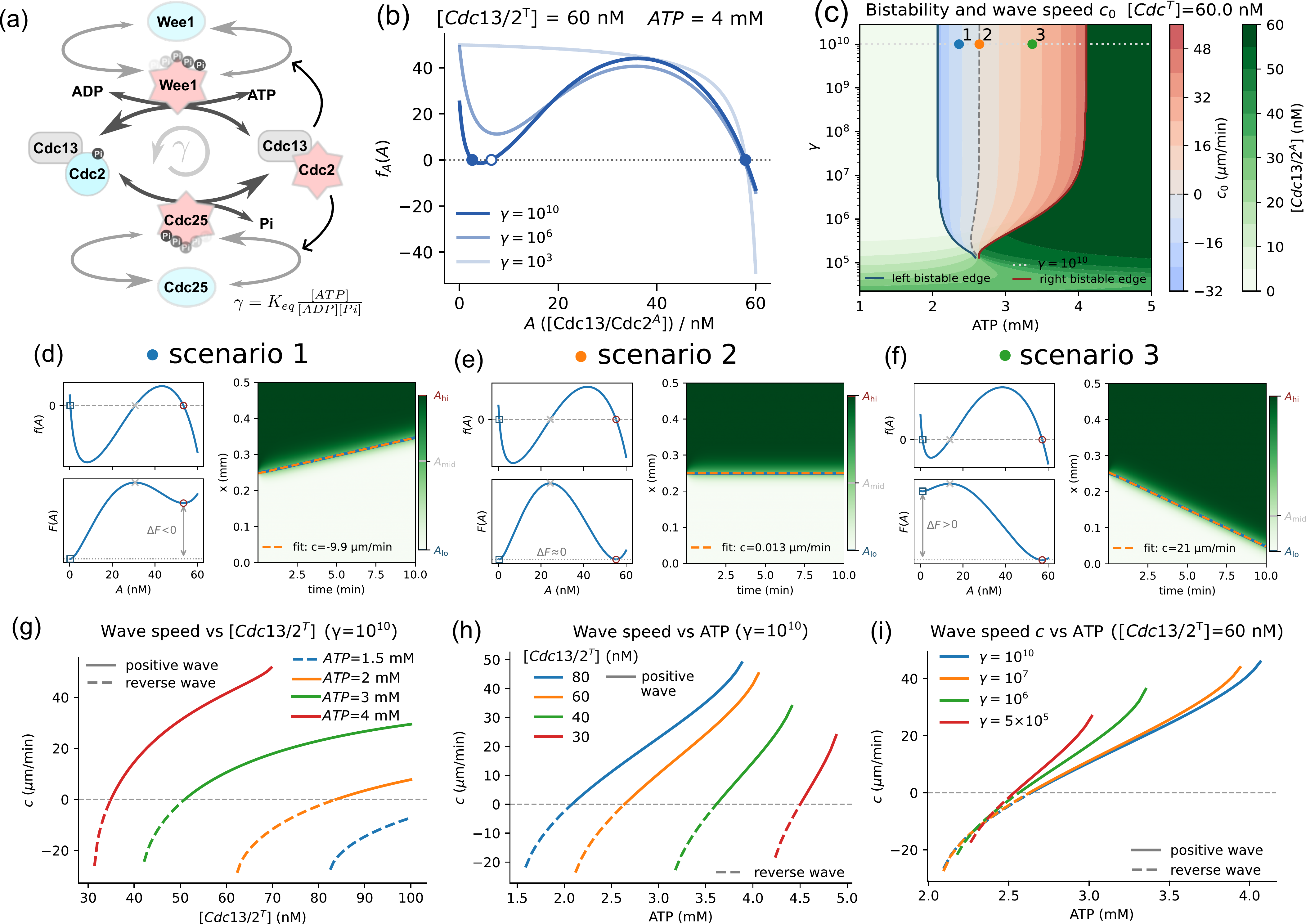}
  \caption{ATP and hydrolysis free energy regulate trigger waves in CDK activation during the G2--M transition.
  (a) Schematic view of the CDK activation circuit.
  (b) $\gamma$-dependence of the Cdk activation reaction term.
  (c) Distribution of bistability and planar wave speed $c_0$.
  Labeled points $1$--$3$ denote representative parameter sets shown in (d--f).
  (d--f) Reaction term, potential function, and propagation profiles for representative cases (blue/purple/red points denote the reverse-wave regime, stationary-front boundary, and forward-wave regime, respectively).
  (g) Dependence of trigger-wave speed on $[\mathrm{Cdc13/Cdc2}^T]$ (fixed $\gamma$).
  (h) Dependence of trigger-wave speed on ATP (fixed $\gamma$).
  (i) Dependence of trigger-wave speed on ATP (fixed $[\mathrm{Cdc13/Cdc2}^T]$).
  }
  \label{fig:g2m_f_and_F}
\end{figure}

Figure~\ref{fig:g2m_f_and_F}(b) shows the reaction term $f(A)$ as a function of $A$. When $f(A)=0$ has three intersections, the system exhibits low- and high-activity stable states separated by an intermediate unstable state. As $\gamma$, which characterizes the free energy of ATP hydrolysis, decreases, the shape of $f(A)$ changes and the intermediate and low fixed points merge and disappear, thereby eliminating bistability. Since $\gamma$ sets the nonequilibrium driving and effective irreversibility of the phosphorylation cycle, it determines whether the bistable structure can be maintained.

As shown in Fig.~\ref{fig:g2m_f_and_F}(c), for fixed $T=60~\mathrm{nM}$, bistability exists only within a finite region in the ATP--$\gamma$ plane. The left and right boundaries correspond to saddle-node bifurcations of the low and high states, respectively. Within the bistable region, the planar wave speed $c_0$ varies substantially with $[\mathrm{ATP}]$. At sufficiently low ATP, reverse waves emerge: the high-activity state no longer invades the low state, but instead retreats as the low state invades the high state. Near physiologically typical levels (indicated by the dashed line $\gamma \approx 10^{10}$), variations in $\gamma$ have a relatively limited influence on both the magnitude and sign of $c_0$, and the wave speed is primarily controlled by $[\mathrm{ATP}]$. However, when $\gamma$ decreases substantially and the system approaches equilibrium, reverse fluxes become stronger and progressively weaken loop directionality; consequently, the bistable region narrows and ultimately disappears. Once bistability is lost, the local switching mechanism vanishes and planar fronts lose the dynamical basis required for sustained trigger-wave propagation. Qualitatively, the dependence of G2--M trigger waves on ATP level and hydrolysis free energy resembles that observed for the S-phase checkpoint Rad53 module, although the detailed ATP--$\gamma$ phase diagram differs quantitatively due to differences in circuit structure\cite{Dianjie_2024}.

Figures~\ref{fig:g2m_f_and_F}(d--f) show representative cases (points $1$--$3$), including the reaction term $f(A)$, the potential function $F(A)$, and the corresponding propagation profiles. When $F(A_{\mathrm{hi}})>F(A_{\mathrm{lo}})$ (i.e., $\Delta F<0$), the low state has lower effective potential and thus invades the high state, yielding a negative wave speed (reverse trigger wave). As ATP increases, $\Delta F$ increases and crosses zero; near $\Delta F=0$ the planar wave speed approaches zero, giving a stationary interface. When the potential wells invert ($\Delta F>0$), the trigger wave becomes forward-propagating.

For fixed $\gamma = 10^{10}$ and given $[\mathrm{ATP}]$, Fig.~\ref{fig:g2m_f_and_F}(g) shows that the wave speed increases monotonically with $T$, and curves at different ATP levels exhibit approximately preserved shape. ATP similarly enhances trigger-wave speed: within the bistable region, for various values of $T$ and $\gamma$, the wave speed increases markedly with $[\mathrm{ATP}]$, as shown in Figs.~\ref{fig:g2m_f_and_F}(h,i). Near physiological $\gamma = 10^{10}$, the wave speed depends only weakly on $\gamma$. Reducing $\gamma$ alone does not substantially change the wave-speed magnitude; instead, it primarily shortens the bistable interval. This behavior differs from the Rad53 system and is clearly visible in the two-dimensional ATP--$\gamma$ phase diagram.

\subsection{G2--M transition with cyclin synthesis/accumulation and APC-mediated cyclin degradation}
\label{subsec:g2m_trigger_adaptive_with_apc}

\subsubsection{Trigger-wave picture with spontaneous accumulation, triggering, and APC-mediated degradation}

\begin{figure}[h!]
  \centering
  \includegraphics[width=1\linewidth]{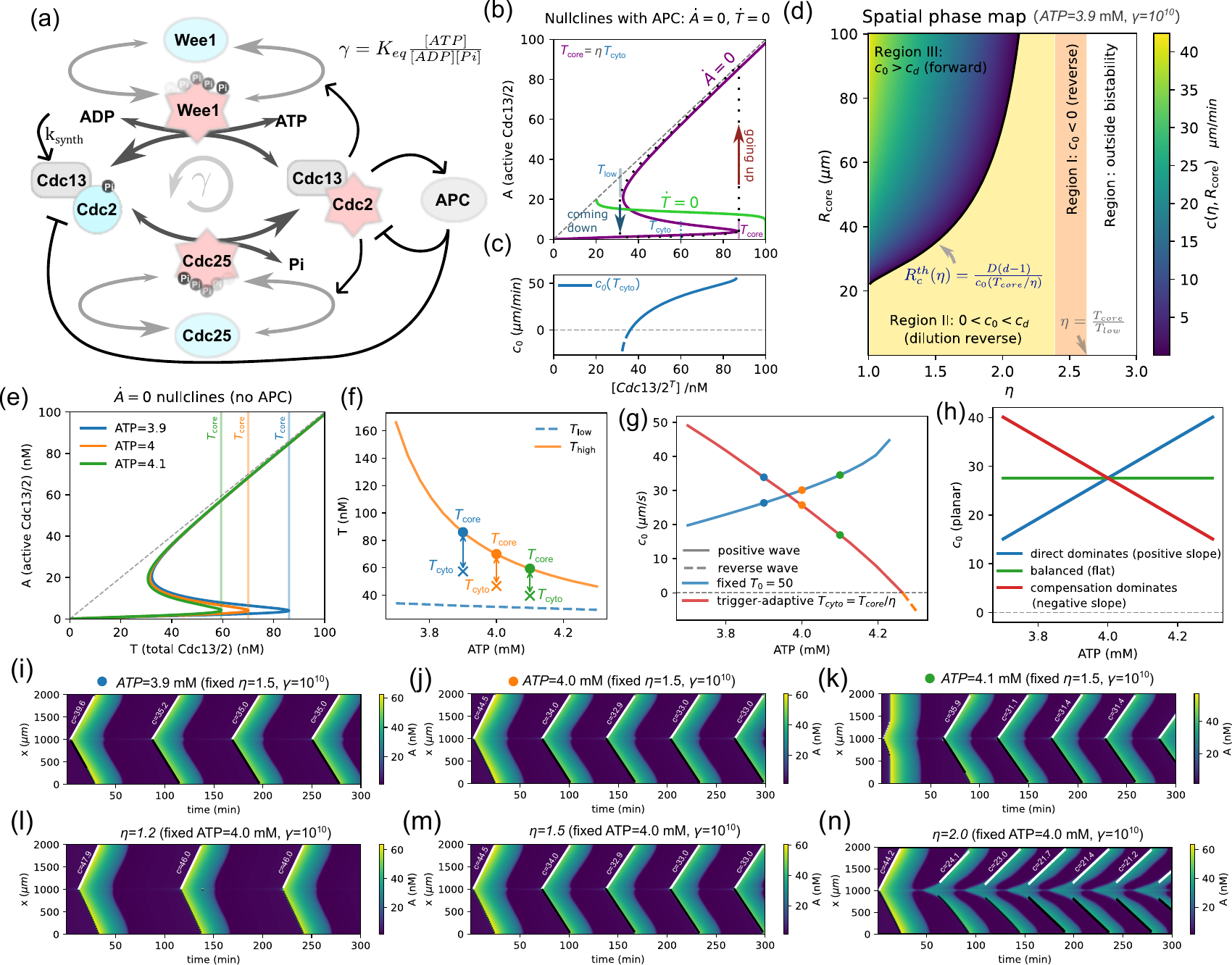}
\caption{Trigger waves in CDK activation with cyclin synthesis and APC-mediated degradation.
(a) Schematic of the Cdk1--APC/C circuit.
(b) Nullclines of the two-variable system: $\dot{A}=0$ (purple) and $\dot{T}=0$ (green). The enriched core activates after crossing the bistable boundary, with $T_{\rm core}\approx T_{\rm high}$.
(c) Cytoplasmic wave speed is set by $T_{\rm cyto}$, with $T_{\rm core}=\eta T_{\rm cyto}$.
(d) At $[\mathrm{ATP}]=3.9~\mathrm{mM}$ and $\gamma=10^{10}$, the $(\eta,R_{\rm core})$ plane is partitioned into a forward-wave region (colored by speed) and three nonpropagating regions: dilution-dominated, reverse-wave, and monostable.
(e--h) ATP has competing effects on wave speed. By lowering the cyclin level required for core activation, increasing ATP reduces $T_{\rm cyto}$ at activation; the resulting ATP dependence of $c_0$ can therefore deviate from, or even reverse relative to, the monotonic increase observed at fixed $T$. Panel (h) summarizes the competition between the direct contribution $\left(\frac{\partial c_0}{\partial [\mathrm{ATP}]}\right)_{T_{\rm cyto}}$ and the indirect contribution $\left(\frac{\partial c_0}{\partial T_{\rm cyto}}\right)_{[\mathrm{ATP}]}\frac{\dd T_{\rm cyto}}{\dd [\mathrm{ATP}]}$.
(i--k) One-dimensional simulations at different ATP levels with $\eta=1.5$ and $\gamma=10^{10}$ fixed.
(l--n) One-dimensional simulations at different enrichment factors with $[\mathrm{ATP}]=4.0$ mM and $\gamma=10^{10}$ fixed.
White lines in (i--n) indicate fitted wave fronts and inferred speeds.
}
  \label{fig:g2m_eta_Rcore_phase}
\end{figure}

In the physiological cell cycle, $[{\rm Cdc13/Cdc2}^T]$ is not constant. It gradually accumulates during G2, and once Cdk becomes activated it subsequently activates APC/C, which induces rapid cyclin degradation, forming a dynamic closed loop of \emph{spontaneous accumulation--triggering--negative-feedback clearance} \cite{novak_design_2008, novak_numerical_1993}. Below we incorporate this biological process into the model and examine the additional phenomenology it introduces for spatial trigger waves. We introduce the dynamics of $T$ and include APC-mediated degradation:
\begin{equation}
\begin{split}
\dot{A}
&= f(A;T,[\mathrm{ATP}],\Delta G)
   - k_{\rm deg}(A)\,A, \\
\dot{T}
&= k_{\rm synth}
   - k_{\rm deg}(A)\,T,
\end{split}
\label{eq:AT_with_APC}
\end{equation}
where $k_{\rm synth}$ is the approximately constant synthesis rate during G2, and $k_{\rm deg}(A)$ is an effective degradation rate that depends on the Cdk activity level $A$. Because APC activation involves additional regulatory complexity and mainly affects the degradation branch of the circuit, we do not model its molecular details explicitly here. Instead, following Yang \textit{et al.} \cite{yangCdk1APCCellCycle2013}, we represent APC-mediated degradation phenomenologically by an ultrasensitive Hill-type function,
\begin{equation}
k_{\rm deg}(A)
= a_{\rm deg} + b_{\rm deg}\,
\frac{1}{1+\left(\dfrac{p_{\rm APC}}{A}\right)^{N_{\rm deg}}},
\label{eq:kdeg_hill}
\end{equation}
so that the model retains the essential CDK self-activation loop together with APC/C-dependent degradation, as schematized in Fig.~\ref{fig:g2m_eta_Rcore_phase}(a).

Following the treatment of Ferrell and coworkers, we assume that spatially there exists a finite region---a \emph{trigger nucleus}---within which Cdc13/Cdk accumulates faster and crosses the activation threshold earlier \cite{changMitoticTriggerWaves2013a}, thereby acting as the initiation site. We characterize this core--cytoplasm difference with a minimal parameter,
\begin{equation}
T_{\rm core} = \eta\,T_{\rm cyto},
\qquad \eta>1,
\label{eq:eta_def}
\end{equation}
where $T_{\rm core}$ denotes the total cyclin--Cdk level in the trigger nucleus, $T_{\rm cyto}$ the cytoplasmic background level, and $\eta$ an enrichment factor arising from phosphorylation-dependent nuclear accumulation of cyclin B/Cdk. In particular, phosphorylation of the cyclin B cytoplasmic retention sequence (CRS) weakens nuclear export and promotes nuclear import, thereby favoring nuclear enrichment of the complex \cite{hagtingTranslocationCyclinB11999,santosSpatialPositiveFeedback2012}. As a result, the enriched core reaches the high-activity state earlier because of its higher $T_{\rm core}$. Once activated, this high-state core acts as a source that launches an outward-propagating trigger wave through reaction--diffusion, thereby driving the surrounding cytoplasm into a coordinated entry into M phase.

Figure~\ref{fig:g2m_eta_Rcore_phase}(b) shows the nullcline structure of the two-variable system after APC-mediated degradation is included. The purple curve denotes the $A$-nullcline, $\dot{A}=0$, which reflects how degradation modifies the effective activation kinetics, whereas the green curve denotes the $T$-nullcline, $\dot{T}=0$, which captures the APC-mediated degradation of $T$. In the $(T,A)$ phase plane, $T$ acts as a relatively slow variable and $A$ as a fast variable. As $T_{\rm core}$ accumulates beyond the bistable boundary, the core undergoes rapid activation and forms a trigger nucleus. At the same time, the relation $T_{\rm core}=\eta T_{\rm cyto}$ links the core level to the cytoplasmic background. The trigger-wave speed $c$ is then determined by the cytoplasmic background level $T_{\rm cyto}$, as summarized by the speed relation in Fig.~\ref{fig:g2m_eta_Rcore_phase}(c). Thus, early activation in the core does not necessarily imply immediate propagation in the cytoplasm, because propagability is ultimately governed by $T_{\rm cyto}$.

Although cyclin accumulation and APC-mediated degradation feedback are included in the reaction circuit, on the timescale of actual trigger-wave propagation the dynamics can be approximated as a single-variable reaction--diffusion process with $T_{\rm cyto}$ treated as constant. We assume a separation of timescales: phosphorylation/dephosphorylation reactions in the Cdk positive-feedback loop occur on a fast timescale, substantially faster than the synthesis-driven accumulation of the total cyclin level $T$. Hence, over the propagation time of the wave front, $T_{\rm cyto}$ can be treated as quasi-static. Moreover, APC/C is activated only by active Cdk, and the degradation process is slower than the rapid phosphorylation dynamics; therefore, during the propagation stage the APC-mediated degradation term can be neglected to leading order.

Under these assumptions, once a trigger wave is formed, its propagation is primarily determined by the cytoplasmic reaction kinetics and diffusion. Accordingly, the planar wave speed can be written as
\begin{equation}
c_0 \;\approx\; c_0\!\left(T_{\rm cyto},[\mathrm{ATP}],\gamma \right)
=c_0(T_{\rm core}/\eta,[\mathrm{ATP}],\gamma ),
\label{eq:c0_cyto}
\end{equation}
where $T_{\rm cyto}$ is the cytoplasmic total cyclin--Cdk complex level.

Accounting jointly for curvature correction and enrichment, Fig.~\ref{fig:g2m_eta_Rcore_phase}(d) shows, at fixed $([\mathrm{ATP}]=3.9~\mathrm{mM},\gamma=10^{10})$, a two-dimensional phase diagram in terms of the enrichment factor $\eta=T_{\rm core}/T_{\rm cyto}$ and the core radius $R_{\rm core}$. This diagram separates propagating and non-propagating regimes, where the latter can be further divided into the dilution-dominated regime, the reverse-wave regime, and the monostable regime:
\begin{enumerate}
  \item \textbf{Propagating regime:}
  Colors indicate the spherical trigger-wave speed. After the trigger nucleus activates, the cytoplasmic background $T_{\rm cyto}$ lies in the forward-propagating bistable region, enabling a sustained outward trigger wave. The propagation speed is then determined by $c_0(T_{\rm cyto},[\mathrm{ATP}],\gamma)$.

  \item \textbf{Dilution-dominated regime:}
  Even if the enriched core ($T_{\rm core}=\eta T_{\rm cyto}$) has crossed the activation threshold and entered the high state, curvature-induced dilution yields $c_0<c_d$, preventing propagation into the cytoplasm.

\item \textbf{Reverse-wave regime:}
Here the planar wave speed determined by $T_{\rm cyto}$ satisfies $c_0<0$. Consequently, the spherical wave speed $c$ remains negative for any core radius. The activated core is therefore nonpropagating.

  \item \textbf{Monostable regime:}
  No bistability exists and thus no trigger wave can form.
\end{enumerate}

The phase diagram shows that core activation is necessary but not sufficient for propagating trigger waves. Notably, being in a non-propagating regime does not necessarily imply collapse of the activated core; rather, the activated core fails to propagate. Only when the cytoplasmic level $T_{\rm cyto}$ becomes sufficiently large---or accumulates over time to cross the threshold so that the system enters the propagating regime---does a trigger wave spread outward from the nucleus. Increasing the initial core radius $R_{\rm core}$ reduces the required accumulation of $T_{\rm cyto}$ and increases the initial propagation speed. Conversely, a smaller core radius enhances spatial dilution, slows the initial front, requires more accumulation of $T_{\rm cyto}$, and may prevent trigger-wave formation altogether.

\subsubsection{Dual effects of ATP on trigger-wave speed}

We find that ATP and $T_{\rm cyto}$ can influence wave speed in an antagonistic manner. This antagonism is illustrated in Figs.~\ref{fig:g2m_eta_Rcore_phase}(e--h). As shown in Fig.~\ref{fig:g2m_eta_Rcore_phase}(e), increasing $[\mathrm{ATP}]$ lowers the triggering threshold of the bistable switch. Consequently, at the triggering time the cytoplasmic level $T_{\rm cyto}=T_{\rm core}/\eta$ also decreases as the threshold decreases [Fig.~\ref{fig:g2m_eta_Rcore_phase}(f)], which can produce an overall trend that is qualitatively opposite to the fixed-$T$ case: although $[\mathrm{ATP}]$ increases, the realized propagation speed $c_0$ may decrease [Fig.~\ref{fig:g2m_eta_Rcore_phase}(g)].

This antagonistic dependence can be expressed via the chain rule,
\begin{equation}
\begin{split}
\frac{\dd c_0}{\dd [\mathrm{ATP}]}
&=
\left(
\frac{\partial c_0}{\partial [\mathrm{ATP}]}
\right)_{T_{\rm cyto}}
+
\left(
\frac{\partial c_0}{\partial T_{\rm cyto}}
\right)_{[\mathrm{ATP}]}
\frac{\dd T_{\rm cyto}}{\dd [\mathrm{ATP}]},
\\[6pt]
T_{\rm cyto}
&=
\frac{T_{\rm core}([\mathrm{ATP}],\gamma)}{\eta}.
\end{split}
\label{eq:chain_rule_comp}
\end{equation}
The first term represents the \emph{direct} effect of ATP: for fixed cytoplasmic background $T_{\rm cyto}$, increasing $[\mathrm{ATP}]$ increases effective reaction rates and thus increases the planar wave speed $c_0(T_{\rm cyto},[\mathrm{ATP}],\gamma)$, consistent with the monotonic ATP dependence in the fixed-$T$ setting. The second term represents an \emph{indirect (compensatory)} effect mediated by the activation threshold: increasing $[\mathrm{ATP}]$ lowers the activation threshold $T_{\rm core}$, implying $\dd T_{\rm cyto}/\dd[\mathrm{ATP}]<0$, while $\partial c_0/\partial T_{\rm cyto}>0$ reflects the monotonic increase of $c_0$ with $T$ at fixed ATP. Depending on parameter combinations, these two contributions jointly determine the overall ATP dependence of the realized wave speed. As summarized in Fig.~\ref{fig:g2m_eta_Rcore_phase}(f), three qualitative behaviors may occur: (i) direct ATP effect dominates and $c_0$ increases with $[\mathrm{ATP}]$; (ii) compensatory effect dominates and $c_0$ decreases with $[\mathrm{ATP}]$; (iii) the two effects balance, leading to weak ATP dependence and near ATP-insensitivity of wave speed.

Notably, in the third case, the coexistence of direct and compensatory effects can make the wave speed relatively insensitive to ATP over a finite range. Conceptually, this resembles the buffering or ``dilution'' effect discussed by Ferrell and coworkers, in which feedback can reduce sensitivity to external perturbations and thereby attenuate the response of trigger-wave dynamics to ATP fluctuations.

\subsubsection{Effects of ATP and $\eta$ on spontaneous repetitive trigger waves}

In spontaneously repeated activation events, we find that both ATP and the enrichment factor $\eta$ affect the activation period. First, increasing $[\mathrm{ATP}]$ lowers the core activation threshold $T_{\rm core}$. Therefore, for a given synthesis rate $k_{\rm synth}$, the G2 accumulation time required to reach triggering shortens,
\begin{equation}
t_{\rm trig}
\;\sim\;
\frac{T_{\rm core}}{k_{\rm synth}}
\quad\Rightarrow\quad
\frac{\dd t_{\rm trig}}{\dd[\mathrm{ATP}]}\;<\;0.
\label{eq:trigger_time_prediction}
\end{equation}

As shown in Figs.~\ref{fig:g2m_eta_Rcore_phase}(i--k), increasing $[\mathrm{ATP}]$ advances the time at which the system activates and initiates a trigger wave, corresponding to a shortened G2 phase. This prediction can be tested directly by modulating cellular ATP levels and measuring G2 duration, e.g., using synchronized populations or single-cell imaging.

Conversely, when $[\mathrm{ATP}]$ is too low, the activation threshold rises substantially, making spontaneous entry into M phase difficult. Dynamically, this corresponds to an energetic gating mechanism: when metabolic driving is insufficient, the G2--M transition is suppressed or even arrested. This regime may be relevant to mitochondrial dysfunction associated with yeast aging phenotypes \cite{ozEngineeringATPImport2025a, li_programmable_2020,shabestaryTrajectoryUnderstandingCellular2025}.

Meanwhile, the enrichment factor $\eta\equiv T_{\rm core}/T_{\rm cyto}$ controls both the timing of triggering and the ability of the activated core to launch a propagating wave, as illustrated in Figs.~\ref{fig:g2m_eta_Rcore_phase}(l--n). Increasing $\eta$ raises the core level $T_{\rm core}$ at a given cytoplasmic background $T_{\rm cyto}$, allowing the core to cross the activation threshold earlier and enter the high state. This tends to shorten G2, but it can also reduce the trigger-wave speed. Consistent with this picture, Maryu and Yang showed in cell-free \textit{Xenopus} extract droplets that nucleocytoplasmic compartmentalization promotes local accumulation and earlier activation of cyclin B1--Cdk1, thereby advancing mitotic entry relative to homogeneous cytoplasmic conditions \cite{maryuNuclearcytoplasmicCompartmentalizationCyclin2022}. If $\eta$ becomes too large, then at the moment of core activation the cytoplasmic level $T_{\rm cyto}$ may still remain outside the propagating regime. In that case, the high-activity state stays temporarily confined near the core rather than expanding immediately into the cytoplasm. Only after $T_{\rm cyto}$ continues to accumulate and enters the propagating regime can a sustained trigger wave develop. In spatiotemporal simulations, this stage appears as a characteristic ``sharp-tip'' feature [Fig.~\ref{fig:g2m_eta_Rcore_phase}(n)]: propagation is initially very slow or transiently stalled after core activation, and becomes sustained only once the cytoplasmic background reaches the propagating range.

Therefore, the time interval between successive trigger waves is jointly determined by the triggering time of the enriched core and the time required for $T_{\rm cyto}$ to reach the propagating regime. When $\eta$ is relatively small, $T_{\rm cyto}$ is already in the propagating regime at the moment of core activation, so a trigger wave is initiated immediately. When $\eta$ is larger, the core may activate earlier, but propagation in the cytoplasm still requires waiting for continued synthesis-driven accumulation of $T_{\rm cyto}$.

Based on our model analysis, we further predict that trigger-wave initiation can be controlled by the initial size of the enriched trigger nucleus. Reducing the nucleus size strengthens curvature effects, thereby delaying activation and, in some cases, preventing the formation of an outward-propagating trigger wave. This tendency may be partially counteracted by increasing ATP levels, since higher ATP lowers the activation threshold and promotes trigger-wave initiation. At the same time, ATP affects wave speed through two competing mechanisms: it accelerates reaction kinetics, but also reduces the threshold at which propagation is launched. Consequently, the wave speed may be relatively insensitive to ATP over a finite range, and ATP variations do not necessarily lead to large changes in propagation speed.

\section{Discussion}

In this work, we first recapitulated the general theoretical description of trigger waves in bistable reaction circuits, including the relation governing wave speed and propagation direction, as well as curvature corrections for fronts in two and three dimensions. We then applied this framework to two representative biological systems containing phosphorylation--dephosphorylation (PdP) cycles---the budding-yeast S-phase DNA damage checkpoint circuit and the fission-yeast G2--M transition circuit---to investigate how ATP concentration and the free energy of ATP hydrolysis regulate trigger-wave dynamics.

Across both systems, ATP level and the ATP-hydrolysis free-energy parameter $\gamma$ regulate trigger-wave behavior by modulating the effective driving and the potential difference $\Delta F$ \cite{qianPhosphorylationEnergyHypothesis2007}. Depending on parameter values, the same circuit can support either forward or reverse propagation. In two and three dimensions, sustained propagation further requires that the initially activated region exceed a critical radius, which is itself jointly controlled by ATP level, hydrolysis free energy, and spatial curvature.

To make the role of energetic driving explicit, we started from biochemical circuits with explicit PdP reactions. In the Rad53 auto-activation circuit in the budding-yeast S-phase checkpoint, which provides a minimal positive-feedback PdP module, the effective dynamics reduce to a cubic-like bistable form \cite{liInterplayATPHydrolysis2024a,jinNonequilibriumStochasticityInfluence2018}, allowing analytical treatment of trigger-wave speed. The analysis predicts that, to leading order, wave speed increases with ATP approximately as a square root, but decreases to zero and becomes negative at sufficiently low ATP, indicating a transition to reverse propagation. The parameter $\gamma$ has a similar qualitative effect, although near physiological conditions its influence is weaker and mainly acts by compressing the bistable region.

Although direct trigger-wave propagation has not yet been demonstrated for the Rad53 circuit, existing experiments support the broader idea that energetic state affects wave propagation. For example, Chen \textit{et al.} found that mitochondrial supplementation increases the speed of apoptotic trigger waves in \textit{Xenopus} egg extracts, with wave speed positively correlated with mitochondrial abundance \cite{chengApoptosisPropagatesCytoplasm2018}. Since the critical trigger-nucleus radius is inversely related to wave speed, reduced ATP also implies a larger critical nucleus size. Because of its relatively simple feedback structure, the Rad53 module may therefore provide a useful minimal platform for testing how ATP level and hydrolysis free energy affect trigger-wave dynamics.

We then turned to the G2--M circuit, where the trigger wave corresponds to a mitotic wave. Although the analysis uses the fission-yeast notation Cdc13/Cdc2, the underlying module represents the conserved mitotic cyclin--CDK control system, corresponding to Cyclin B/Cdk1 in vertebrates. In the simplified setting without Cdc13 synthesis or APC-mediated degradation, numerical results show trends qualitatively similar to those in the Rad53 system: increasing ATP increases wave speed, whereas sufficiently low ATP can drive the system into a reverse-propagation regime. In this circuit, however, $\gamma$ primarily regulates the extent of the bistable region rather than the wave speed itself, and its influence remains weak near physiological values. The total mitotic cyclin level, denoted here by Cdc13, also acts as an effective control parameter that shifts propagation thresholds and tunes the wave speed.

This analysis also highlights the reverse-propagation regime, which has received relatively little attention. In the trigger-wave picture, the speed scale follows a Luther-type relation, whereas the propagation direction is set by the sign of $\Delta F$. In the G2--M system, lowering either ATP or the total mitotic cyclin level can reverse the sign of $\Delta F$ and thus induce reverse propagation. Experimentally, however, this regime would require tight control of the system, likely including suppression of APC-mediated degradation and external control of cyclin level.

To make the G2--M analysis more biologically realistic, we next incorporated Cdc13 synthesis/accumulation and APC-mediated degradation, and introduced the concept of a trigger nucleus to describe spontaneous local activation. In this setting, the nucleus is assumed to be enriched in Cdc13/Cdc2 relative to the surrounding cytoplasm, with the enrichment measured by a factor $\eta$. Larger $\eta$ allows the core to cross the activation threshold earlier, but whether the activated state can propagate outward still depends on the cytoplasmic background level and on the spatial scale of the trigger nucleus. As a result, early local activation does not necessarily imply immediate wave propagation: if the effective wave speed at the time of activation is still negative, the system remains in a localized activated state until the cytoplasmic background accumulates sufficiently to support propagation.

With synthesis and degradation included, ATP dependence becomes more subtle. ATP not only accelerates reaction kinetics, tending to increase wave speed, but also lowers the triggering threshold, which can reduce the cytoplasmic background level at the time of activation. These two effects act antagonistically, so the ATP dependence of wave speed need not be monotonic and may become partially buffered over a finite range. This provides one possible interpretation of experimentally observed robustness. ATP-dependent shifts in the CDK activation threshold have been observed in microdroplet systems \cite{guanRobustTunableMitotic2018}, and trigger-wave speeds in \textit{Xenopus} cytoplasmic extracts have recently been reported to be robust against cytoplasmic dilution and concentration changes \cite{huangRobustTriggerWave2024}. Although the latter work offers a detailed mechanistic account, ATP variations coupled to cytoplasmic concentration may also contribute to this apparent robustness.

The critical-radius prediction can be related to several observations in mitotic trigger-wave experiments. In \textit{Xenopus} egg extracts, the measured wave speed is about $60\,\mu\text{m}/\text{min}$. Using a three-dimensional diffusion coefficient of $D\approx 600\,\mu\text{m}^2/\text{min}$, our critical-radius relation gives an estimated trigger-nucleus radius of about $20\,\mu\text{m}$. This is comparable to the typical size of the \textit{Xenopus} nucleus\cite{jevticNuclearSizeScaling2015,jevticElucidatingNuclearSize2018,heijoDNAContentContributes2020,wesleyOrganelleSizeScaling2020}, consistent with experiments showing that mitotic trigger waves originate from the nucleus rather than the centrosome \cite{noletNucleiDetermineSpatial2020,afanzarNucleusServesPacemaker2020}. Moreover, Nolet \textit{et al.} found that reducing nuclear size below a threshold delays mitotic entry and can prevent the emergence of mitotic trigger waves \cite{noletNucleiDetermineSpatial2020}, in agreement with the existence of a minimal trigger-nucleus radius.

The enrichment factor $\eta$ may also be tested indirectly through nucleocytoplasmic organization. For a given nucleus, increasing the associated cytoplasmic volume may enhance the effective local enrichment generated by nuclear import. In \textit{Drosophila} embryos, Hayden \textit{et al.} introduced nonuniform nuclear distributions and observed that mitotic waves preferentially initiate from the side with lower nuclear density \cite{haydenManipulatingNatureEmbryonic2022}. This is consistent with the model prediction that larger effective enrichment facilitates earlier threshold crossing and wave initiation. Together with the nuclear-origin and nuclear-size observations above, these results support the roles of critical trigger radius and local enrichment in trigger-wave formation.

Several limitations should be noted. First, because some key physiological parameters are not yet sufficiently constrained, our conclusions are mainly qualitative or semi-quantitative. In particular, in the G2--M system the ATP dependence of wave speed arises from two competing effects---changes in reaction rates and shifts in activation thresholds---so its quantitative form remains uncertain and will need to be resolved experimentally.Second, reverse trigger waves are unlikely to occur under normal physiological conditions, because they are incompatible with the usual temporal direction of cell-cycle progression. Experimentally, their realization would likely require stable maintenance of a high-activity state together with suppression of APC-mediated degradation and external control of Cdc13 levels. Third, although a Rad53-like PdP positive-feedback circuit may provide a simple platform for testing energetic effects, trigger-wave behavior has not yet been directly established in such a system.
Finally, our analysis focuses on trigger-wave mechanisms, whereas real cell-cycle dynamics may involve both trigger waves and sweep waves, with possible crossover between the two \cite{haydenManipulatingNatureEmbryonic2022, vergassolaMitoticWavesEarly2018, taliaWavesEmbryonicDevelopment2022a}. Spatial heterogeneity, time-dependent parameters, and intracellular structural complexity may all influence mode selection, and these effects remain outside the scope of the present model.

In summary, we developed a unified trigger-wave framework for PdP-based biochemical systems and used it to analyze how ATP concentration and ATP-hydrolysis free energy regulate propagation direction, wave speed, and the critical trigger-nucleus scale. Applying this framework to the Rad53 checkpoint circuit and the G2--M transition circuit shows how energetic driving shapes trigger-wave behavior in cell-cycle regulation. More broadly, this work provides a quantitative basis for understanding how ATP level and hydrolysis free-energy constraints influence cell-scale spatiotemporal signal propagation in biological trigger-wave systems \cite{denekeChemicalWavesCell2018, gelensSpatialTriggerWaves2014}.

\appendix

\section{Front analysis and wave-speed expression of trigger waves}
\label{appendix:wave_speed_derivation}

The theory of propagating fronts in bistable reaction--diffusion systems has been analyzed in detail in the literature. 
Accordingly, our purpose here is not to provide a comprehensive review, but rather to collect the main results used in the present work in a compact form, so as to facilitate reading and interpretation of the trigger-wave picture adopted in the main text. 
In particular, the direction of front propagation and the general properties of front and pulse solutions are discussed in the ``Front and pulse propagation'' subsection of ``Patterns in Chemical Reactions'' in Cross and Hohenberg's review~\cite{crossPatternFormationOutside1993}, while curvature corrections in higher dimensions are treated in the ``Higher-dimensional patterns'' subsection of the same review. 
In addition, Murray's \emph{Mathematical Biology} discusses exact traveling-wave solutions for cubic bistable kinetics in the section ``Biological Waves: Single Species Models''~\cite{murray_mathematical_1989}; this cubic form, and certain extensions of it, is often referred to as the reduced Nagumo equation and is closely related to the FitzHugh--Nagumo model.

We consider a one-dimensional reaction--diffusion system and introduce the traveling-wave coordinate 
$z = x - ct$, with $u(x,t) = U(z)$. 
Substituting this ansatz into Eq.~\eqref{eq:RD_general} yields the ordinary differential equation for the traveling-wave profile
\begin{equation}
D \frac{\dd^2 U}{\dd z^2} + c \frac{\dd U}{\dd z} + f(U)=0 .
\label{eq:TW_ode}
\end{equation}

We focus on monotonic wave fronts connecting two stable fixed points,
which satisfy the boundary conditions
\begin{equation}
\lim_{z\to-\infty}U(z)=u_{\mathrm{hi}},
\qquad
\lim_{z\to+\infty}U(z)=u_{\mathrm{lo}},
\qquad
\lim_{z\to\pm\infty}U'(z)=0 .
\label{eq:TW_bc}
\end{equation}

\paragraph{Step 1: Multiplying by $U'$ and integrating.}

Multiplying Eq.~\eqref{eq:TW_ode} by $U'(z)$ gives
\begin{equation}
D U''U' + c (U')^2 + f(U)U' = 0 .
\label{eq:TW_mulUp}
\end{equation}

Integrating over $z\in(-\infty,+\infty)$ yields
\begin{equation}
\int_{-\infty}^{+\infty} D U''U' \,\dd z
+
c\int_{-\infty}^{+\infty} (U')^2\,\dd z
+
\int_{-\infty}^{+\infty} f(U)\,U'\,\dd z
=0 .
\label{eq:TW_int_all}
\end{equation}

\paragraph{Step 2: Contribution of the diffusion term.}

Noting that $U''U'=\frac{1}{2}\frac{\dd}{\dd z}(U')^2$, we obtain
\begin{equation}
\int_{-\infty}^{+\infty} D U''U' \,\dd z
=
\frac{D}{2}(U')^2\Big|_{-\infty}^{+\infty}
=0 ,
\label{eq:TW_int_diff_zero}
\end{equation}
where the boundary condition $U'(\pm\infty)=0$ has been used.

\paragraph{Step 3: Reaction term and definition of a potential function.}

For the reaction term we use the substitution $\dd U = U'\dd z$, which gives
\begin{equation}
\int_{-\infty}^{+\infty} f(U)\,U'\,\dd z
=
\int_{u_{\mathrm{hi}}}^{u_{\mathrm{lo}}} f(u)\,\dd u
=
-\int_{u_{\mathrm{lo}}}^{u_{\mathrm{hi}}} f(u)\,\dd u .
\label{eq:TW_int_reaction}
\end{equation}

To facilitate interpretation, we introduce an effective potential function $F(u;\theta)$ defined by
\begin{equation}
\frac{\partial F}{\partial u}(u;\theta) = -f(u;\theta),
\label{eq:potential_def}
\end{equation}
where $\theta$ denotes the set of parameters appearing in the reaction term.
Using this definition, we obtain
\begin{equation}
\int_{u_{\mathrm{lo}}}^{u_{\mathrm{hi}}} f(u;\theta)\,\dd u
=
-\int_{u_{\mathrm{lo}}}^{u_{\mathrm{hi}}} \frac{\partial F}{\partial u}(u;\theta)\,\dd u
=
F(u_{\mathrm{lo}};\theta)-F(u_{\mathrm{hi}};\theta).
\label{eq:int_f_to_F}
\end{equation}

\paragraph{Step 4: Potential difference and wave-speed identity.}

Substituting Eqs.~\eqref{eq:TW_int_diff_zero} and \eqref{eq:TW_int_reaction}
into Eq.~\eqref{eq:TW_int_all} yields
\begin{equation}
c\int_{-\infty}^{+\infty} (U')^2\,\dd z
=
\int_{u_{\mathrm{lo}}}^{u_{\mathrm{hi}}} f(u;\theta)\,\dd u .
\label{eq:c_integral_identity}
\end{equation}

We define the effective potential difference between the two stable states as
\begin{equation}
\Delta F(\theta)
\equiv
F(u_{\mathrm{lo}};\theta)-F(u_{\mathrm{hi}};\theta)
\equiv
\int_{u_{\mathrm{lo}}}^{u_{\mathrm{hi}}} f(u;\theta)\,\dd u .
\label{eq:DeltaF_def}
\end{equation}

This quantity characterizes the effective driving force for the transition between the two states.
When $\Delta F>0$, the high-activity state becomes globally favored;
when $\Delta F<0$, the low-activity state is preferred;
and $\Delta F=0$ corresponds to the Maxwell point at which the two states have equal effective potential.

The physical meaning of $\Delta F$ does not originate from equilibrium thermodynamics,
but rather from the asymmetry of reaction fluxes under nonequilibrium driving conditions.

Equation~\eqref{eq:c_integral_identity} can therefore be written as
\begin{equation}
c\int_{-\infty}^{+\infty} (U')^2\,\dd z = \Delta F(\theta).
\label{eq:c_DeltaF_identity}
\end{equation}

because the additive constant in the potential $F(u)$ cancels in the difference,
$\Delta F$ is uniquely defined.
The general expression for the wave speed therefore follows as
\begin{equation}
c = \frac{\Delta F(\theta)}{\displaystyle \int_{-\infty}^{+\infty} (U')^2\,\dd z},
\label{eq:c_ratio_form_appendix}
\end{equation}
where $U'(z)\equiv \dd U/\dd z$ denotes the derivative of the traveling-wave profile with respect to $z$.
In the main text, to avoid introducing the additional traveling-wave notation $U(z)$, we write the same denominator in the equivalent form
$\int_{-\infty}^{+\infty} \left(\dd u/\dd z\right)^2 \dd z$.

\subsection{Curvature correction and the eikonal relation in the thin-front limit}
\label{appendix:curvature_correction}

In two- or three-dimensional systems, trigger waves generally propagate as curved interfaces. 
When the intrinsic front thickness $\ell$ is much smaller than the local radius of curvature $R$, 
the interface can be treated using the thin-front approximation. 
In this limit the front may be regarded as a geometrical interface whose normal motion is governed by the local reaction--diffusion dynamics.

Consider the reaction--diffusion equation
\begin{equation}
\frac{\partial u}{\partial t}=D\nabla^2 u+f(u).
\label{eq:RD_general_appendix}
\end{equation}

For radially symmetric configurations, let $u(\mathbf{x},t)=u(r,t)$ with $r=|\mathbf{x}|$. 
In $d$ spatial dimensions the Laplacian becomes
\begin{equation}
\frac{\partial u}{\partial t}
=
D\left(
\frac{\partial^2 u}{\partial r^2}
+
\frac{d-1}{r}\frac{\partial u}{\partial r}
\right)
+
f(u).
\label{eq:RD_radial}
\end{equation}

Let the position of the propagating front be $r=R(t)$ and introduce the local coordinate
\begin{equation}
z=r-R(t),
\end{equation}
which measures distance from the moving interface.
Under the thin-front approximation, the solution near the interface can be expressed as
\begin{equation}
u(r,t)\approx U(z),
\end{equation}
where $U(z)$ represents the stationary wave profile in the comoving frame.

Substituting this ansatz into Eq.~\eqref{eq:RD_radial} gives
\begin{equation}
-\dot R\,U'
=
D\left(
U''+\frac{d-1}{R+z}U'
\right)
+
f(U).
\label{eq:front_equation}
\end{equation}

Assuming that the front is localized within a narrow region $|z|\ll R$, 
the curvature term can be approximated as
\begin{equation}
\frac{1}{R+z}\approx\frac{1}{R}.
\end{equation}

Equation~\eqref{eq:front_equation} then reduces to
\begin{equation}
D U''+\left(\dot R-\frac{D(d-1)}{R}\right)U'+f(U)=0.
\label{eq:front_curvature}
\end{equation}

This equation has the same structure as the planar traveling-wave equation
\begin{equation}
D U''+c_0 U'+f(U)=0,
\label{eq:planar_wave}
\end{equation}
provided that the normal velocity of the interface satisfies
\begin{equation}
\frac{\dd R}{\dd t}
=
c_0-\frac{D(d-1)}{R}.
\label{eq:Rdot_curvature}
\end{equation}

Equation~\eqref{eq:Rdot_curvature} shows that the normal propagation speed of the front equals the planar wave speed $c_0$, corrected by a curvature-dependent term. 
Writing the curvature of the interface as
\begin{equation}
\kappa=\frac{d-1}{R},
\end{equation}
this result takes the well-known eikonal form
\begin{equation}
v_n=c_0-D\kappa,
\end{equation}
where $v_n$ denotes the normal velocity of the propagating interface. 
This relation states that local curvature modifies the propagation speed through diffusive relaxation along the interface.

\subsection{Critical nucleus radius and geometric threshold for trigger-wave initiation}

The curvature-corrected propagation law \eqref{eq:Rdot_curvature} reveals an important property of trigger waves: 
their initiation depends not only on the local reaction kinetics but also on the spatial scale of the activated region.

Setting $\dd R/\dd t=0$ in Eq.~\eqref{eq:Rdot_curvature} yields the critical nucleus radius
\begin{equation}
R_c=\frac{D(d-1)}{c_0}.
\label{eq:Rc_general}
\end{equation}

This expression reflects a competition between two effects. 
The planar wave speed $c_0$ characterizes the intrinsic ability of the reaction kinetics to drive the interface outward, whereas the curvature term $D(d-1)/R$ represents the diffusive tendency of a curved interface to contract.

If the initial activated region has radius
\begin{equation}
R_0>R_c,
\end{equation}
then $\dd R/\dd t>0$ and the nucleus expands, eventually developing into a self-sustained trigger wave. 
Conversely, if
\begin{equation}
R_0<R_c,
\end{equation}
then $\dd R/\dd t<0$ and the activated region shrinks due to curvature-driven diffusion.

Therefore the radius $R_c$ defines a geometric threshold separating expanding trigger events from perturbations that decay. 
Only perturbations exceeding this spatial scale are able to overcome diffusive smoothing and grow into a propagating trigger wave.

\subsection{Explicit wave-speed formula under a cubic approximation}

The simplest polynomial capable of producing bistability is a cubic nonlinearity. 
For reaction terms exhibiting two stable fixed points, the dynamics in the bistable regime can often be approximated by the factorized cubic form
\begin{equation}
f(u)
=
-a\,(u-r_{\mathrm{lo}})(u-r_{\mathrm{mid}})(u-r_{\mathrm{hi}}),
\qquad
r_{\mathrm{lo}}<r_{\mathrm{mid}}<r_{\mathrm{hi}},
\qquad
a>0 .
\label{eq:cubic_factor_form}
\end{equation}

\paragraph{Potential difference.}

Substituting Eq.~\eqref{eq:cubic_factor_form} into the definition of $\Delta F$ yields
\begin{equation}
\Delta F
=
-a\int_{r_{\mathrm{lo}}}^{r_{\mathrm{hi}}}
(u-r_{\mathrm{lo}})(u-r_{\mathrm{mid}})(u-r_{\mathrm{hi}})
\,\dd u .
\end{equation}

Evaluating the integral gives
\begin{equation}
\Delta F
=
\frac{a}{12}(r_{\mathrm{hi}}-r_{\mathrm{lo}})^3
\left(
r_{\mathrm{lo}}+r_{\mathrm{hi}}-2r_{\mathrm{mid}}
\right).
\label{eq:DeltaF_roots}
\end{equation}

Thus the sign of $\Delta F$ is determined solely by the combination
\(
r_{\mathrm{lo}}+r_{\mathrm{hi}}-2r_{\mathrm{mid}}.
\)

\paragraph{Exact traveling-wave solution.}

For cubic Nagumo-type reaction terms, the traveling-wave equation admits an exact monotonic solution. 
The front profile can be written in logistic form
\begin{equation}
U(z)=r_{\mathrm{lo}}+\frac{r_{\mathrm{hi}}-r_{\mathrm{lo}}}{1+\exp(\lambda z)},
\label{eq:logistic_front}
\end{equation}
which is equivalent to a $\tanh$ representation.

Substituting this profile into the traveling-wave equation determines the front steepness parameter
\begin{equation}
\lambda
=
(r_{\mathrm{hi}}-r_{\mathrm{lo}})\sqrt{\frac{a}{2D}} .
\label{eq:lambda_relation}
\end{equation}

\paragraph{Explicit wave-speed formula.}

Using the general identity
\(
c_0=\Delta F/\int (U')^2\,\dd z
\)
together with the analytic front profile \eqref{eq:logistic_front}, one obtains the explicit planar wave speed
\begin{equation}
c_0
=
\sqrt{\frac{Da}{2}}
\left(
r_{\mathrm{lo}}+r_{\mathrm{hi}}-2r_{\mathrm{mid}}
\right).
\label{eq:c0_nagumo}
\end{equation}

\paragraph{Relation between $c_0$ and $\Delta F$.}

Comparing Eqs.~\eqref{eq:c0_nagumo} and \eqref{eq:DeltaF_roots} shows that both the wave speed $c_0$ and the potential difference $\Delta F$ are controlled by the same combination of roots,
\(
r_{\mathrm{lo}}+r_{\mathrm{hi}}-2r_{\mathrm{mid}}.
\)
Consequently, the direction of propagation and the sign of the effective driving force are determined by the relative position of the unstable root $r_{\mathrm{mid}}$ with respect to the two stable states.

\section{Rad53 reaction network and ATP-dependent kinetics}
\label{app:Rad53_thermo_derivation}

Rad53 activation is driven jointly by the upstream DNA-damage signal $S$ and the positive feedback of phosphorylated Rad53. 
Following the reaction scheme shown in Fig.~\ref{fig:Rad53_scheme}, we define four reaction fluxes corresponding to forward and reverse phosphorylation cycles
\begin{align}
J_1^+ &= (k_1 S + k_2 R_p^2)\,R\,[ATP], \\
J_2^+ &= k_-\,R_p, \\
J_1^- &= (k_1' S + k_2' R_p^2)\,R_p\,[ADP], \\
J_2^- &= k_-'\,R\,[P_i] .
\end{align}

The net production rate of phosphorylated Rad53 is therefore
\begin{equation}
\frac{\dd R_p}{\dd t}
=
(J_1^+ + J_2^-)-(J_2^+ + J_1^-).
\label{eq:Rp_flux_balance_app}
\end{equation}

Using conservation of the total Rad53 concentration and substituting the flux expressions into Eq.~\eqref{eq:Rp_flux_balance_app}, one obtains a mass-action kinetic equation for $R_p$. 
Introducing the rescaled time $\tau=k_- t$ and collecting parameters into dimensionless groups leads to a reduced dynamical equation.

Following the formulation of Jin \emph{et al.}~\cite{jinNonequilibriumStochasticityInfluence2018}, the dynamics can be written in nondimensional form as
\begin{equation}
\frac{\dd u}{\dd \tau}
=
(p+q u^2)(1-u)-(1+\beta)u,
\qquad
u \equiv R_p \in [0,1].
\label{eq:rad53_nondim}
\end{equation}

Here $p$ and $q$ characterize the strength of the upstream signal and the nonlinear positive feedback, respectively, while $\beta$ represents the contribution of reverse reactions.

\subsection{Explicit dependence on ATP and the nonequilibrium parameter}

To distinguish the roles of ATP concentration and phosphorylation free energy, the parameters in Eq.~\eqref{eq:rad53_nondim} can be written explicitly in terms of $[ATP]$ and the nonequilibrium driving parameter
\begin{equation}
\gamma=\frac{k_1 k_2}{k_1' k_2'}\,
\frac{[ATP]}{[ADP][P_i]},
\qquad
\Delta G = RT \ln \gamma .
\end{equation}

After eliminating $[ADP]$ using the thermodynamic relation, the kinetic parameters take the form
\begin{equation}
p=\frac{1}{\alpha}\left(1+\xi_0[ATP]\right),
\qquad
q=\frac{\eta_0}{\alpha}[ATP],
\label{eq:pq_atp}
\end{equation}
with
\begin{equation}
k_\gamma \equiv \frac{[P_i]}{K_{\mathrm{eq}}},
\qquad
\xi_0 \equiv \frac{k_1' S}{k_-\,k_\gamma},
\qquad
\eta_0 \equiv \frac{k_2'}{k_-\,k_\gamma},
\label{eq:xi0_eta0_def}
\end{equation}
where $K_{\mathrm{eq}}$ is the equilibrium constant for ATP synthesis. 
The quantity $\xi_0$ depends explicitly on the DNA-damage signal strength $S$, whereas $\eta_0$ is determined by $[P_i]$ together with the reaction-rate constants. 
Under the common assumption that $[P_i]$ remains approximately constant, $k_\gamma$ and hence $\eta_0$ can be treated as constants independent of the metabolic state. 
Thus the nonlinear activation parameters $p$ and $q$ depend explicitly only on ATP concentration, whereas the nonequilibrium parameter $\gamma$ enters through the relative weighting of forward and reverse reaction fluxes.

Substituting these relations into the reaction term yields an explicit decomposition
\begin{equation}
f_{\mathrm{full}}(u)
=
f_{\infty}(u)-\varepsilon\,g(u),
\qquad
\varepsilon=\frac{1}{\gamma},
\end{equation}
with
\begin{align}
f_{\infty}(u)
&=
\frac{1}{\alpha}(1-u)
+
\frac{[ATP]}{\alpha}(\xi_0+\eta_0 u^2)(1-u)
-
u, \\
g(u)
&=
[ATP](\xi_0+\eta_0 u^2)u .
\end{align}

Here $f_{\infty}(u)$ corresponds to the far-from-equilibrium limit $\gamma\to\infty$, while the term proportional to $\varepsilon$ represents the weakening of the nonlinear activation due to reverse reaction fluxes.

\subsection{ATP dependence of the wave speed and critical nucleus size}
\label{app:Rad53_ATP_wave_speed_derivation}

Within the bistable regime, the Rad53 reaction term in Eq.~\eqref{eq:rad53_nondim} can be locally approximated by a cubic form,
\begin{equation}
f(u)\approx -a\,(u-r_{\mathrm{lo}})(u-r_{\mathrm{mid}})(u-r_{\mathrm{hi}}),
\end{equation}
with $r_{\mathrm{lo}}<r_{\mathrm{mid}}<r_{\mathrm{hi}}$. 

For the Rad53 network, comparison of Eq.~\eqref{eq:rad53_nondim} with the cubic normal form shows that the effective cubic coefficient can be approximated as $a \sim k_- \frac{\eta_0}{\alpha}[ATP]$. Substituting this relation into Eq.~\eqref{eq:c0_nagumo} yields
\begin{equation}
c_0
\sim
\sqrt{
\frac{Dk_-}{2}\,
\frac{\eta_0}{\alpha}\,
[ATP]
}
\left(
r_{\mathrm{lo}}+r_{\mathrm{hi}}-2r_{\mathrm{mid}}
\right).
\label{eq:c0_atp_scaling}
\end{equation}

Thus, provided that the root combination
\(
r_{\mathrm{lo}}+r_{\mathrm{hi}}-2r_{\mathrm{mid}}
\)
does not vary strongly across the bistable regime, the dominant scaling is
\begin{equation}
c_0 \propto \sqrt{[ATP]}.
\end{equation}

Using the curvature-corrected threshold relation
\begin{equation}
R_c=\frac{D(d-1)}{c_0},
\end{equation}
the corresponding critical nucleus radius becomes
\begin{equation}
R_c
\sim
(d-1)\,
\sqrt{
\frac{2D}{k_-}\,
\frac{\alpha}{\eta_0}\,
\frac{1}{[ATP]}
}\,
\frac{1}{
r_{\mathrm{lo}}+r_{\mathrm{hi}}-2r_{\mathrm{mid}}
}.
\label{eq:Rc_atp_scaling}
\end{equation}
Accordingly,
\begin{equation}
R_c \propto [ATP]^{-1/2},
\end{equation}
under the same approximation.

These relations predict that increasing ATP concentration accelerates trigger-wave propagation while simultaneously lowering the minimum nucleus size required for sustained expansion.

\section{Thermodynamically consistent model for the Cdk1 activation circuit}
\label{appendix:g2m_model}

To investigate how ATP concentration and phosphorylation free energy influence the G2--M transition, we employ a thermodynamically consistent model for the Cdc13/Cdc2 activation network based on the well-established Cdk1--Wee1--Cdc25 regulatory circuit. This model is based on the reversible phosphorylation framework developed by Zhao \emph{et al.}~\cite{zhaoNonequilibriumNonlinearKinetics2016}. As the detailed derivation is available in that work, we provide here only the essential model structure and the final kinetic expressions required for the present analysis.

The core process is the phosphorylation--dephosphorylation cycle of the Cdc13/Cdc2 complex. Wee1 catalyzes the inhibitory phosphorylation of Cdc13/Cdc2, whereas Cdc25 removes this phosphate group to activate the kinase. The elementary reactions can be written as
\begin{equation}
\small
Cdc13/Cdc2^{A} + Wee1^{A} + ATP
\;\xrightleftharpoons[k_{-1}]{k_{+1}}\;
Cdc13/Cdc2^{A}\!\cdot Wee1^{A}\!\cdot ATP
\;\xrightleftharpoons[k_{-2}]{k_{+2}}\;
Cdc13/Cdc2^{In} + Wee1^{A} + ADP ,
\end{equation}

\begin{equation}
Cdc13/Cdc2^{In} + Cdc25^{A}
\;\xrightleftharpoons[k_{-3}]{k_{+3}}\;
Cdc13/Cdc2^{In}\!\cdot Cdc25^{A}
\;\xrightleftharpoons[k_{-4}]{k_{+4}}\;
Cdc13/Cdc2^{A} + Cdc25^{A} + P_i .
\end{equation}

Here $A$ and $In$ denote the active and inactive states of the Cdc13/Cdc2 complex. The rate constants $k_{\pm i}$ describe the forward and reverse microscopic reaction steps.

Because the cellular concentration of Cdc2 is much larger than that of Cdc13 and their binding affinity is high, the total Cdc13/Cdc2 complex concentration can be approximated as
\begin{equation}
[Cdc13/Cdc2^{T}]
=
[Cdc13/Cdc2^{A}]
+
[Cdc13/Cdc2^{In}] .
\end{equation}

The dynamics of the active Cdc13/Cdc2 complex is therefore determined by the balance between the dephosphorylation flux mediated by Cdc25 and the phosphorylation flux mediated by Wee1,
\begin{equation}
\frac{d[Cdc13/Cdc2^{A}]}{dt}
=
J_{P}
-
J_{W}.
\end{equation}

Following the quasi–steady-state approximation for the enzyme–substrate complexes, the effective rate constants can be written as
\begin{equation}
a_{+}=\frac{k_{+1}k_{+2}}{k_{-1}+k_{+2}},
\qquad
a_{-}=\frac{k_{-1}k_{-2}}{k_{-1}+k_{+2}},
\end{equation}

\begin{equation}
b_{+}=\frac{k_{+3}k_{+4}}{k_{-3}+k_{+4}},
\qquad
b_{-}=\frac{k_{-3}k_{-4}}{k_{-3}+k_{+4}} .
\end{equation}

The phosphorylation and dephosphorylation fluxes are then given by
\begin{equation}
J_W=
\frac{
\left\{
a_{+}[ATP][Cdc13/Cdc2^{A}]
-
a_{-}[ADP]
([Cdc13/Cdc2^{T}]-[Cdc13/Cdc2^{A}])
\right\}
[Wee1^{A}]
}{
1+
\frac{a_{+}[ATP][Cdc13/Cdc2^{A}]}{k_{+2}}
+
\frac{a_{-}[ADP]([Cdc13/Cdc2^{T}]-[Cdc13/Cdc2^{A}])}{k_{-1}}
},
\end{equation}

\begin{equation}
J_P=
\frac{
\left\{
b_{+}([Cdc13/Cdc2^{T}]-[Cdc13/Cdc2^{A}])
-
b_{-}[P_i][Cdc13/Cdc2^{A}]
\right\}
[Cdc25^{A}]
}{
1+
\frac{b_{+}([Cdc13/Cdc2^{T}]-[Cdc13/Cdc2^{A}])}{k_{+4}}
+
\frac{b_{-}[P_i][Cdc13/Cdc2^{A}]}{k_{-3}}
}.
\end{equation}

Unlike traditional cell-cycle models that treat phosphorylation cycles as irreversible processes, this formulation explicitly retains both forward and reverse reactions in order to maintain thermodynamic consistency. In this framework, the nonequilibrium driving force is characterized by the phosphorylation free energy
\begin{equation}
\Delta G = RT\ln\gamma ,
\end{equation}
where the nonequilibrium parameter $\gamma$ is defined as
\begin{equation}
\gamma
=
\frac{a_{+}b_{+}[ATP]}
{a_{-}b_{-}[ADP][P_i]} .
\end{equation}

The parameter $\gamma$ measures the distance from thermodynamic equilibrium of the phosphorylation–dephosphorylation cycle. Increasing ATP concentration or decreasing ADP and $P_i$ enhances $\gamma$, thereby strengthening the nonequilibrium driving force of the Cdk activation circuit.

\subsection{Steady-state relations for Wee1 and Cdc25 activation}

The activities of Wee1 and Cdc25 are regulated through multisite phosphorylation
mediated by Cdc13/Cdc2. Following the thermodynamically consistent
phosphorylation--dephosphorylation framework developed by
Zhao \emph{et al.}~\cite{zhaoNonequilibriumNonlinearKinetics2016},
the multisite modification cycle can be described by a sequence of
phosphorylation states whose steady-state distribution is determined
by the ratio of forward and reverse modification rates.

Under the commonly used sequential phosphorylation mechanism and
assuming that the microscopic reaction constants are independent of
the phosphorylation state, the steady-state fractions of the modified
states form a geometric sequence controlled by the parameter
\begin{equation}
\sigma = \frac{\alpha}{\beta},
\end{equation}
where $\alpha$ and $\beta$ denote the effective phosphorylation and
dephosphorylation rates, respectively.

As a result, the steady-state activities of Wee1 and Cdc25 can be
expressed as functions of the active Cdc13/Cdc2 level. The fractions
of active Wee1 and Cdc25 take the forms
\begin{equation}
[Wee1^{A}]
=
\rho\,[Wee1^{T}]
\left(
1-
\frac{\sigma_1^{M}-\sigma_1^{M+1}}
{1-\sigma_1^{M+1}}
\right),
\label{eq:wee1_active}
\end{equation}

\begin{equation}
[Cdc25^{A}]
=
\lambda\,[Cdc25^{T}]
\frac{\sigma_2^{N}-\sigma_2^{N+1}}
{1-\sigma_2^{N+1}},
\label{eq:cdc25_active}
\end{equation}
where $M$ and $N$ denote the numbers of phosphorylation sites of Wee1
and Cdc25, respectively. The coefficients $\rho$ and $\lambda$ represent
the fractions of free enzyme available for modification.

The parameters $\sigma_1$ and $\sigma_2$ depend on the active Cdc13/Cdc2
concentration and the nonequilibrium driving force $\gamma$ through
\begin{equation}
\sigma_1
=
\frac{
\gamma\,\delta_1(\tilde c/\tilde d)+\tilde c
}{
\delta_1+\tilde d
},
\qquad
\sigma_2
=
\frac{
\gamma\,\delta_2(\tilde c/\tilde d)+\tilde c
}{
\delta_2+\tilde d
},
\end{equation}
with
\begin{equation}
\delta_1=\frac{[Cdc13/Cdc2^{A}]}{[P_{\mathrm{Wee1}}]},
\qquad
\delta_2=\frac{[Cdc13/Cdc2^{A}]+[kinase^{Cdc25}]}{[P_{\mathrm{Cdc25}}]} .
\end{equation}
and
\begin{equation}
\tilde c=\frac{b_{-}[P_i]}{a_{-}[ADP]},
\qquad
\tilde d=\frac{b_{+}}{a_{-}[ADP]} .
\end{equation}

Here $\gamma$ characterizes the nonequilibrium phosphorylation free
energy associated with ATP hydrolysis.
Under the commonly used assumption that the enzyme--substrate
dissociation constants are large ($L,Q\gg1$), the free-enzyme fractions
satisfy $\rho\approx1$ and $\lambda\approx1$.
Therefore, Eqs.~\eqref{eq:wee1_active} and \eqref{eq:cdc25_active}
provide closed functional relations linking the activities of Wee1
and Cdc25 to the active Cdc13/Cdc2 concentration.

\begin{table}[H]
\centering
\caption{Parameters used in the Rad53 model.}
\begin{tabular}{llll}
\hline
Parameter & Meaning & Value & Unit \\
\hline
$\alpha$ & Reaction scaling parameter & $10^{7}$ & dimensionless \\

$\eta_0$ & Positive feedback coefficient & $2\times10^{7}$ & mM$^{-1}$ \\

$\xi_0$ & Basal activation coefficient & $6.32\times10^{4}$ & mM$^{-1}$ \\

$[ATP]$ & ATP concentration & $4$ & mM \\

$\gamma$ & Nonequilibrium parameter & $10^{10}$ & dimensionless \\
\hline
\end{tabular}
\label{tab:rad53_parameters}
\end{table}

All other parameters are absorbed into dimensionless combinations
through nondimensionalization and therefore do not appear explicitly.

\begin{table}[H]
\centering
\caption{Parameters used in the thermodynamically consistent G2--M transition model.}
\begin{tabular}{llll}
\hline
Parameter & Meaning & Value & Unit \\
\hline
$T$ & Total Cdc13/Cdc2 concentration & $60.0$ & nM \\

$\gamma$ & Nonequilibrium parameter & $10^{10}$ & dimensionless \\

$[ATP]$ & ATP concentration & $4.0$ & mM \\

$[P_i]$ & Inorganic phosphate concentration & $1.0$ & mM \\

$K_{\mathrm{eq}}$ & Equilibrium constant for ATP hydrolysis & $4.9\times10^{8}$ & mM$^{-1}$ \\

\hline
$[Wee1]^T$ & Total Wee1 concentration & $50.0$ & nM \\

$[Cdc25]^T$ & Total Cdc25 concentration & $100.0$ & nM \\

\hline
$a_{+}$ & Forward phosphorylation coefficient & $1.6\times10^{-1}$ & nM$^{-2}$\,min$^{-1}$ \\

$a_{-}$ & Reverse phosphorylation coefficient & $1.0\times10^{-5}$ & nM$^{-2}$\,min$^{-1}$ \\

$b_{+}$ & Forward dephosphorylation coefficient & $1.0\times10^{-1}$ & nM$^{-1}$\,min$^{-1}$ \\

$b_{-}$ & Reverse dephosphorylation coefficient & $3.27\times10^{-6}$ & nM$^{-2}$\,min$^{-1}$ \\

$k_{-1}$ & Reverse microscopic rate constant & $1.0$ & min$^{-1}$ \\

$k_{+2}$ & Forward microscopic rate constant & $1.0$ & min$^{-1}$ \\

$k_{-3}$ & Reverse microscopic rate constant & $1.0$ & min$^{-1}$ \\

$k_{+4}$ & Forward microscopic rate constant & $2.0$ & min$^{-1}$ \\

\hline
$p_{\mathrm{Wee1}}$ & Feedback-shaping parameter for Wee1 & $100.0$ & nM \\

$p_{\mathrm{Cdc25}}$ & Feedback-shaping parameter for Cdc25 & $160.0$ & nM \\

$k_{\mathrm{Cdc25}}$ & Basal kinase input assisting Cdc25 activation & $25.0$ & nM \\

$M$ & Number of phosphorylation sites on Wee1 & $5$ & dimensionless \\

$N$ & Number of phosphorylation sites on Cdc25 & $5$ & dimensionless \\

\hline
$a_{\mathrm{deg}}$ & Basal cyclin degradation rate & $0.01$ & min$^{-1}$ \\

$b_{\mathrm{deg}}$ & APC-dependent cyclin degradation coefficient & $0.04$ & min$^{-1}$ \\

$k_{\mathrm{synth}}$ & Cyclin synthesis rate & $1.0$ & min$^{-1}$ \\

$p_{\mathrm{APC}}$ & APC activation threshold parameter & $15$ & nM \\

$N_{\mathrm{deg}}$ & Hill coefficient for APC-dependent degradation & $17$ & dimensionless \\
\hline
\end{tabular}
\label{tab:g2m_parameters}
\end{table}

The reverse coefficient $b_{-}$ is not chosen independently, but is fixed by the thermodynamic consistency condition
\[
\frac{a_{+}b_{+}}{b_{-}a_{-}} = K_{\mathrm{eq}},
\]
which ensures the correct equilibrium constraint for the ATP-coupled phosphorylation cycle.

\bibliography{reference}

\end{document}